\documentclass{article}

\usepackage{PRIMEarxiv}

\usepackage[utf8]{inputenc} 
\usepackage[T1]{fontenc}    
\usepackage{url}            
\usepackage{booktabs}       
\usepackage{amsfonts}       
\usepackage{nicefrac}       
\usepackage{microtype}      
\usepackage{lipsum}
\usepackage{fancyhdr}       
\usepackage{graphicx}       

\usepackage{utils/packages}

\definecolor{color_rebut}{RGB}{255, 140, 0}

\colorlet{gray}{black!75}
\colorlet{shadecolor}{gray!80}
\definecolor{uc3mblue}{RGB}{0, 14, 120}

\definecolor{matlab_blue}{HTML}{0072BD}
\definecolor{matlab_orange}{HTML}{D95319}
\definecolor{matlab_yellow}{HTML}{EDB120}
\definecolor{matlab_purple}{HTML}{7E2F8E}
\definecolor{matlab_green}{HTML}{77AC30}
\definecolor{matlab_cyan}{HTML}{4DBEEE}
\definecolor{matlab_red}{HTML}{A2142F}

\definecolor{training_color}{HTML}{434E78}
\definecolor{testing_color}{HTML}{E97F4A}

\definecolor{diff_color}{HTML}{C5705D}
\definecolor{diff_color_original}{HTML}{B68B82}
\definecolor{ae_color}{HTML}{708871}

\definecolor{triangle_color}{HTML}{5D79C5}
\definecolor{triangle_color_original}{HTML}{7E73E2}

\newacronym{DDPM}{DDPM}{Denoising Diffusion Probabilistic Model}
\newacronym{DDIM}{DDIM}{Denoising Diffusion Implicit Model}
\newacronym{CFD}{CFD}{Computational Fluid Dynamics}
\newacronym{ML}{ML}{Machine Learning}
\newacronym{CRM}{CRM}{Common Research Model}
\newacronym{Cp}{$C_p$}{Pressure Coefficient}
\newacronym{RANS}{RANS}{Reynolds-Averaged Navier-Stokes}
\newacronym{M}{$\mathrm{M}$}{Mach number}
\newacronym{AoA}{$\alpha$}{Angle of Attack}
\newacronym{Inail}{$\delta_1$}{Inboard Aileron Deflection Angle}
\newacronym{Outail}{$\delta_2$}{Outboard Aileron Deflection Angle}
\newacronym{Elev}{$\delta_3$}{Elevator Deflection Angle}
\newacronym{Htp}{$\delta_4$}{Horizontal Tailplane Deflection Angle}
\newacronym{PCA}{PCA}{Principal Component Analysis}
\newacronym{POD}{POD}{Proper Orthogonal Decomposition}
\newacronym{NN}{NN}{Neural Network}
\newacronym{MLP}{MLP}{Multi-Layer Perceptron}
\newacronym{AE}{AE}{Autoencoder}
\newacronym{b-VAE}{$\beta$-VAE}{$\beta$-Variational Autoencoder}
\newacronym{GAN}{GAN}{Generative Adversarial Network}
\newacronym{GPR}{GPR}{Gaussian Process Regression}
\newacronym{MAE}{MAE}{Mean Absolute Error}
\newacronym{MSE}{MSE}{Mean Squared Error}
\newacronym{SiLU}{SiLU}{Sigmoid Linear Unit}
\newacronym{LR}{LR}{Learning Rate}
\newacronym{CH}{CH}{Convex Hull}
\newacronym{RMSE}{RMSE}{Root Mean Squared Error}
\newacronym{LRI}{LRI}{Local Reliability Index}
\newacronym{GRI}{GRI}{Global Reliability Index}

\graphicspath{{figures/}}     

\title{Signal-Aware Conditional Diffusion Surrogates for Transonic Wing Pressure Prediction}

\author{
  V\'ictor Franc\'es-Belda \\
  Theoretical and Computational Aerodynamics Group \\ Flight Physics Department. Subdirectorate General  for Aeronautical Systems\\ 
  Spanish National Institute for Aerospace Technology (INTA)\\ 
  \vspace{-0.2cm}\\
  Department of Aerospace Engineering \\
  Universidad Carlos III de Madrid \\
  Madrid, Spain \\
  \texttt{vfrabel@inta.es} \\
  \And
  Carlos Sanmiguel Vila\\
  Aerial Platforms Department \\ Subdirectorate General for Aeronautical Systems\\ 
  Spanish National Institute for Aerospace Technology (INTA)\\
  \vspace{-0.2cm}\\
  Department of Aerospace Engineering\\
  Universidad Carlos III de Madrid \\
  Madrid, Spain \\
  \texttt{csanvil@inta.es} \\
  \And
  Rodrigo Castellanos \\
  Department of Aerospace Engineering \\
  Universidad Carlos III de Madrid \\
  Madrid, Spain \\
  \texttt{rcastell@ing.uc3m.es} \\
}

\begin{document}
\maketitle
\begin{abstract}
Accurate and efficient surrogate models for aerodynamic surface pressure fields are essential for accelerating aircraft design and analysis, yet deterministic regressors trained with pointwise losses often smooth sharp nonlinear features. This work presents a conditional denoising diffusion probabilistic model for predicting surface pressure distributions on the NASA Common Research Model wing under varying conditions of Mach number, angle of attack, and four control surface deflections. The framework operates on unstructured surface data through a principal component representation used as a non-truncated, reversible linear reparameterization of the pressure field, enabling a fully connected architecture. A signal-aware training objective is derived by propagating a reconstruction loss through the diffusion process, yielding a timestep-dependent weighting that improves fidelity in regions with strong pressure gradients. The stochastic sampling process is analyzed through repeated conditional generations, and two diagnostic metrics are introduced, the Local Reliability Index and Global Reliability Index, to relate sampling-induced spread to reconstruction error. Relative to the considered deterministic baselines, the proposed formulation reduces mean absolute error and improves the reconstruction of suction peaks, shock structures, and control surface discontinuities. The sampling-induced spread exhibits strong correspondence with surrogate error, supporting its interpretation as a qualitative reliability indicator rather than calibrated uncertainty quantification.
\end{abstract}


\newpage
\section*{Nomenclature}
{\renewcommand\arraystretch{1.05}
\noindent\begin{longtable*}{@{}l @{\quad=\quad} p{0.72\textwidth}@{}}
$b$                & design space dimension \\
$\mathbf{C}$       & design matrix \\
$C_p$              & pressure coefficient vector\\
$\mathbf{c}$       & conditioning vector (flight condition sextuplet) \\
$c$                & chord, m \\
$\mathbf{D}$       & database's matrix \\
$d$                & number of dataset's samples \\
$\mathcal{F}$      & PCA projection operator \\
$\mathcal{F}^{-1}$ & inverse PCA projection operator \\
$\mathcal{G}$      & layer projection operator \\
$\mathcal{G}^{-1}$ & inverse layer projection operator \\
$h$                & number of mesh points \\
$k$                & embedding dimension \\
$l$                & number of testing flight conditions \\
M                  & Mach number \\
$m$                & number of retained PCA components \\
$\mathcal{N}$      & Gaussian distribution \\
$N$                & ensemble size \\
$n$                & number of flight conditions \\
$r$                & feature-space dimensionality \\
$\mathbf{T}$       & time matrix \\
$T$                & total number of timesteps \\
$t$                & diffusion timestep \\
$\mathbf{X}$       & dataset of pressure fields \\
$\mathbf{x}$       & physical $C_p$ sample \\
$\mathbf{x}^*$     & predicted $C_p$ sample \\
$\mathbf{Z}$       & latent data matrix \\
$\mathbf{z}$       & latent representation of $\mathbf{x}$ \\
$\mathbf{z}^*$     & latent representation of $\mathbf{x}^*$ \\
$\mathbf{z}_0$     & clean, uncorrupted signal \\
$\mathbf{z}_t$     & noisy signal at timestep $t$ \\
$\alpha$         & angle of attack, deg \\
$\beta_t$        & variance schedule parameter at timestep $t$ \\
$\gamma_t$       & complementary variance schedule parameter at timestep $t$,  $\gamma_t = 1 - \beta_t$ \\
$\overline{\gamma}_t$ & cumulative product of $\gamma_t$, $\overline{\gamma}_t = \prod_{i=1}^t \gamma_i$ \\
$\delta_i$       & deflection angle of $i$-th surface, deg \\
$\boldsymbol{\varepsilon}$      & Gaussian noise \\
$\boldsymbol{\varepsilon}_\theta$ & noise predicted by the model \\
$\eta$                            & span percentage from wing's root \\
$\theta$                          & parameters of a neural network \\
$\boldsymbol{\mu}_\theta$         & ensemble's mean pressure field \\
$\boldsymbol{\sigma}_\theta$      & ensemble's standard deviation \\
$\rho$                            & Pearson coefficient \\
$\omega$                          & fraction of sorted mesh points based on model's deviation


\end{longtable*}}

\twocolumn 
\section{Introduction}\label{s:introduction}
\gls{CFD} simulations, together with wind tunnel campaigns and flight testing, constitute the main sources of high-quality aerodynamic data for design optimization, manufacturing, and performance analysis in the aerospace industry \cite{he2017aerodatafusion, anhichem2022multi}. Although experimental approaches generally provide the highest fidelity, their widespread use is limited by cost, operational complexity, and the effort required for data acquisition and processing \cite{malik2012roleof}. \gls{CFD} offers a more flexible and cost-effective alternative, but high-fidelity simulations remain computationally demanding, particularly for complex flow topologies. To alleviate the cost of generating aerodynamic databases, \gls{ML} techniques have been increasingly adopted as surrogate models, enabling rapid predictions from pre-computed datasets without repeatedly solving the governing equations \cite{yondo2019review, du2021rapid, du2022airfoil}.

A central difficulty in aerodynamic surrogate modeling is that the quantities of interest are often high-dimensional fields containing sharp nonlinear features. Modern deterministic surrogates span a wide spectrum of representations and architectures, including graph \glspl{NN} for aircraft surface pressure prediction \cite{hines2023graph}, mesh-based neural fields (implicit neural representations) \cite{catalani2024neuralfields, catalani2026towards}, and neural operator frameworks designed to learn solution operators to accelerate parametric and shape-dependent simulations \cite{shukla2024deepoperators, hu2025pano}. While these approaches can be highly accurate, they are typically deployed as point-estimate predictors and frequently optimized with pointwise reconstruction losses such as \gls{MSE}. As a consequence, the learned mapping is statistically driven toward conditional-average predictions. In aerodynamics, this behavior becomes especially problematic in transonic regimes: if the surrogate exhibits spatial ambiguity in the shock position, the \gls{MSE}-optimal prediction tends to smear the pressure jump into a smooth ramp rather than preserving a sharp discontinuity. This degradation is not merely a visual artifact: a smeared shock prediction directly corrupts the integrated wave drag and pitching moment, quantities that are central to transonic aircraft design and performance assessment.

To address these limitations, generative modeling frameworks have recently attracted growing attention in fluid mechanics. Early efforts employed \glspl{GAN}, either directly or in combination with \glspl{AE}, for the synthesis of aerodynamic flow fields and pressure distributions \cite{nandal2023synergistic, wang2023cgan}. However, the instability and mode-collapse issues commonly associated with \glspl{GAN} have motivated the exploration of diffusion-based generative models. In particular, diffusion probabilistic models \cite{dickstein2015noneq, ho2020ddpm, nichol2021improved, song2020score} provide a robust likelihood-free training strategy and can represent richer conditional distributions than deterministic regressors \cite{nichol2021beatgan, saharia2021iterative}. This representational advantage has been recognized across a growing body of work in fluid mechanics and aerodynamics.

In aerodynamics, \glspl{DDPM} have been investigated for uncertainty-aware airfoil-flow prediction \cite{liu2024uncertainty}, while diffusion transformers have been proposed for Reynolds-averaged airfoil simulations \cite{xiang2025aerodit, ogbuagu2026foildiff}, and transformer-guided diffusion models have been employed for flow-field reconstruction and super-resolution \cite{liu2025flowvitdiff}. Diffusion formulations have also been explored beyond direct field prediction, for example, through latent-space samplers for aerodynamic shape generation and optimization \cite{wei2024diffairfoil}. More broadly in fluid mechanics, latent diffusion with neural field embeddings has been employed for stochastic spatiotemporal turbulence generation in three-dimensional domains \cite{du2024turbulence}, while graph-based and point-wise diffusion formulations have extended these models to irregular meshes, strong geometric variability, and large-scale physical systems \cite{lino2025diffusiongraph, kim2025pointwise}. Together, these studies show that diffusion models are evolving from image-oriented generative tools into a flexible surrogate family for fluid mechanics applications. This evolution spans structured and unstructured discretizations, deterministic and stochastic prediction settings, and both physical-field reconstruction and design-oriented generation. While accelerated deterministic samplers can reduce computational cost \cite{song2022ddim}, the stochastic sampling process remains an intrinsic aspect of diffusion models and provides access to ensemble statistics.

Within this context, the present work investigates a complementary perspective: a conditional \gls{DDPM}-based regression surrogate for predicting high-fidelity aerodynamic surface pressure distributions on the three-dimensional wing of a transport aircraft configuration in transonic conditions. The proposed framework operates in a \gls{PCA}-based modal space, where \gls{PCA} is used as a full, non-truncated, reversible linear reparameterization of the pressure field rather than as a reduced-order model. This enables a fully-connected denoising network acting on modal coefficients, thereby accommodating unstructured surface data without requiring a structured-grid representation. In addition, this work examines the implications of the standard diffusion noise-matching objective for aerodynamic pressure prediction in regions with strong gradients, and derives a signal-aware reformulation with a timestep-dependent weighting designed to improve fidelity in such regions.

Beyond prediction accuracy, this study analyzes the stochastic response of the diffusion model under repeated sampling. Unlike approaches that interpret diffusion-sampling variability as calibrated predictive uncertainty \cite{liu2024uncertainty, ogbuagu2026foildiff, xu2025diffpcno}, the present work does not pursue quantitative uncertainty quantification. Instead, the variability induced by the generative process is examined as a qualitative sensitivity and reliability indicator. To this end, two diagnostic metrics are introduced: the \gls{LRI}, which measures local correspondence between predictive spread and reconstruction error, and the \gls{GRI}, which provides the analogous relation at the global-condition level. The objective is therefore not to claim calibrated uncertainty estimates, but to assess whether the intrinsic stochasticity of the diffusion model can identify regions and operating conditions associated with larger surrogate errors.

The remainder of this article is organized as follows: \autoref{s:theoretical_background} introduces the foundational concepts of \glspl{DDPM}. Section \ref{s:database} presents the \gls{CFD} database and test case. Section \ref{s:methodology} describes the surrogate framework, including the architecture of the model, the conditioning process, and the proposed signal-aware objective. The results of this methodology are outlined in \autoref{s:results}, involving a comparative analysis against several baselines and a sensitivity analysis. Finally, \autoref{s:conclusion} highlights the findings of the work.

\section{Theoretical background}\label{s:theoretical_background}
\glspl{DDPM} \cite{ho2020ddpm} are a class of latent variable models inspired by non-equilibrium thermodynamics \cite{dickstein2015noneq}, originally developed for image generation \cite{ho2020ddpm, nichol2021beatgan, saharia2021iterative}, and subsequently extended to fluid-mechanics and aerodynamic applications \cite{du2024turbulence, wei2024diffairfoil}. The essential idea of these models is to progressively corrupt a signal into pure Gaussian noise through an iterative forward diffusion process, and then learn a reverse diffusion process that restores structure from the corrupted state, yielding a highly flexible and tractable generative model of the data. While deterministic samplers can accelerate inference \cite{song2022ddim}, the standard Markovian \gls{DDPM} sampling strategy is retained in this work to maximize prediction fidelity and preserve the stochasticity required for variance quantification.

The forward diffusion process slowly adds Gaussian noise to the original signal over $T$ timesteps, following a Markov chain \cite{kuntz2020markov}. The amount of noise injected at a given timestep $t \in [1, 2, \ldots, T]$ is controlled by a diffusion parameter $\beta_t$, commonly referred to as the variance schedule. In general, the data distribution $\mathbf{z}$ at $t$, given the data distribution at $t-1$, can be expressed as
\begin{equation}\label{eq:forward_process}
   \mathbf{z}_t = \sqrt{1-\beta_t}\,\mathbf{z}_{t-1} + \sqrt{\beta_t}\,\boldsymbol{\varepsilon}, \quad \boldsymbol{\varepsilon} \sim \mathcal{N}(\mathbf{0}, \mathbf{I}),
\end{equation}
where $\boldsymbol{\varepsilon}$ is sampled from a multivariate standard Gaussian distribution of mean $\mathbf{0}$ and covariance $\mathbf{I}$. Here, $\mathbf{z}_0$ denotes the uncorrupted sample, whereas $\mathbf{z}_T$ represents the end of the forward process, where the signal is completely destroyed. Note that $\beta_t$  may vary with $t$ depending on the selected noise schedule \cite{guo2025noise}. As $t$ approaches $T$, $\mathbf{z}_t$ converges to a Gaussian noise distribution. However, $\mathbf{z}_t$ can also be sampled directly from $\mathbf{z}_0$ in closed form, without iterating through the intermediate steps:
\begin{equation}\label{eq:forward_process_alphabar}
    \mathbf{z}_t = \sqrt{\overline{\gamma}_t}\,\mathbf{z}_0 + \sqrt{1 - \overline{\gamma}_t}\,\boldsymbol{\varepsilon}, 
\end{equation}
with
\begin{align*}
    \gamma_t = 1 - \beta_t,\\
    \overline{\gamma}_t = \prod_{i=1}^t \gamma_i.
\end{align*}

Note that the parameter $\gamma$ is generally referred to as $\alpha$ in the diffusive literature. In the present work, the notation is changed to avoid confusion with the angle of attack, which is denoted by $\alpha$.

The reverse diffusion process is likewise defined in a Markovian manner. While the forward process progressively corrupts the clean signal $\mathbf{z}_0$, the reverse process aims to restore it. Specifically, the objective is to sample from $q(\mathbf{z}_{t-1}\mid\mathbf{z}_t)$, which denotes the distribution of the cleaner signal $\mathbf{z}_{t-1}$ conditioned on the noisy observation $\mathbf{z}_t$. Under the Gaussian forward noising process, this reverse conditional can be written as
\begin{equation}\label{eq:true_posterior}
    q(\mathbf{z}_{t-1}\mid\mathbf{z}_t) = \mathcal{N} (\mathbf{z}_{t-1}; \boldsymbol{\mu}, \boldsymbol{\Sigma}), 
\end{equation}
where $\boldsymbol{\mu}$ and $\boldsymbol{\Sigma}$ denote the mean vector and covariance matrix of the distribution. However, computing the exact values of $\boldsymbol{\mu}$ and $\boldsymbol{\Sigma}$ requires marginalizing over the entire data distribution $q(\mathbf{z}_0)$, which is intractable. In simple terms, perfectly reversing the noising process would require complete knowledge of the probability density of every possible uncorrupted signal, which is unavailable. Because $q$ cannot be computed exactly, it is approximated by a parametrized distribution $p_\theta$ defined through a \gls{NN}, such that  
\begin{equation}\label{eq:approx_posterior} 
    p_\theta(\mathbf{z}_{t-1}\mid\mathbf{z}_t) = \mathcal{N}(\mathbf{z}_{t-1}; \boldsymbol{\mu}_\theta(\mathbf{z}_t, t), \boldsymbol{\Sigma}_\theta(\mathbf{z}_t, t)).
\end{equation}

To fully specify $p_\theta$, the \gls{NN} must define its mean $\boldsymbol{\mu}_\theta$ and covariance $\boldsymbol{\Sigma}_\theta$. In the standard, simplified \gls{DDPM} formulation, the covariance is fixed from the forward process parameters as 
\begin{equation}
    \mathbf{\Sigma}_\theta(\mathbf{z}_t, t) = \beta_t \mathbf{I},
\end{equation}
thereby reducing the learning task to the prediction of  $\boldsymbol{\mu}_\theta$. This leads to the following parameterization \cite{ho2020ddpm}:
\begin{equation}\label{eq:mean_parametrization}
    \boldsymbol{\mu}_\theta(\mathbf{z}_t, t) = \dfrac{1}{\sqrt{\gamma_t}} \left( \mathbf{z}_t - \dfrac{\beta_t}{\sqrt{1 - \overline{\gamma}_t}} \boldsymbol{\varepsilon}_\theta(\mathbf{z}_t, t)\right),
\end{equation}
where $\boldsymbol{\varepsilon}_\theta$ is the function approximator used to predict the added noise $\boldsymbol{\varepsilon}$ from $\mathbf{z}_t$. Consequently, $\mathbf{z}_{t-1}$ can be sampled from $p_\theta$ as
\begin{equation}\label{eq:reverse_process}
    \mathbf{z}_{t-1} = \dfrac{1}{\sqrt{\gamma_t}} \left( \mathbf{z}_t - \dfrac{\beta_t}{\sqrt{1 - \overline{\gamma}_t}} \boldsymbol{\varepsilon}_\theta(\mathbf{z}_t, t)\right) + \sqrt{\beta_t}\, \mathbf{s},
\end{equation}
with $\mathbf{s} \sim \mathcal{N}(\mathbf{0}, \mathbf{I})$ if $t > 1$, and $\mathbf{s} = \mathbf{0}$ otherwise, so that the final denoised signal is not re-corrupted.

Therefore, once $\boldsymbol{\Sigma}_\theta$ is fixed and $\boldsymbol{\mu}_\theta$ is parameterized through $\boldsymbol{\varepsilon}_\theta$, the learning problem reduces to predicting  the noise added to $\mathbf{z}_t$ in order to reconstruct $\mathbf{z}_{t-1}$. In practice, this predictor is often implemented with a U-Net-type architecture \cite{ronneberger2015unet}. Under this formulation, the standard training objective minimizes the \gls{MSE} between the actual and predicted noise \cite{ho2020ddpm}, namely
\begin{equation}\label{eq:training_objective}
    \mathcal{L}(\theta) = \mathbb{E}_{\mathbf{z}_0, \boldsymbol{\varepsilon}, t} \left[ \lVert \boldsymbol{\varepsilon} - \boldsymbol{\varepsilon}_\theta (\mathbf{z}_t, t)\rVert_2^2 \right].
\end{equation}

\subsection{Training and sampling algorithms}\label{ss:training_sampling}
The training procedure follows the forward diffusion process explained above. A set of input samples is provided to the denoising network, and each sample is corrupted at a randomly selected timestep according to  \autoref{eq:forward_process_alphabar}. The network is then trained to predict the noise that has been added to each corrupted sample. Once training is completed, the function $\boldsymbol{\varepsilon_\theta}(\mathbf{z}_t, t)$ is fully defined.

The sampling procedure starts from pure Gaussian noise. The reverse diffusion process is then applied iteratively, following \autoref{eq:reverse_process}, so as to progressively recover a new uncorrupted sample. At each timestep, the trained predictor $\boldsymbol{\varepsilon_\theta}(\mathbf{z}_t, t)$ is used to estimate the noise component associated with the current corrupted signal.

\subsection{Conditioning the diffusion process}\label{ss:conditioning}
The standard \gls{DDPM} framework described above models the marginal distribution of the data $q(\mathbf{z}_0)$, allowing the generation of samples that, while not present in the dataset, are consistent with it. In many practical applications, however, it is desirable to control the generation process according to a specific set of attributes or labels, for example, a specific flight condition \cite{zhan2024survey}. This leads to conditional diffusion models, which aim to represent the conditional distribution $q(\mathbf{z}_0 \mid \mathbf{c})$, where $\mathbf{c}$ denotes the conditioning information (e.g., class labels, text embeddings, or physical parameters). In this setting, the reverse diffusion process is modified so that it depends on $\mathbf{c}$ at each timestep, namely 
\begin{equation}
    p_\theta(\mathbf{z}_{t-1} \mid \mathbf{z}_t, \mathbf{c}) = \mathcal{N}(\mathbf{z}_{t-1}; \boldsymbol{\mu}_\theta(\mathbf{z}_t, t, \mathbf{c}), \beta_t \mathbf{I}),
\end{equation}
where the difference with \autoref{eq:approx_posterior} is the explicit dependence on $\mathbf{c}$. As described above, $\boldsymbol{\mu}_\theta$ is parameterized following \autoref{eq:mean_parametrization}, but the noise predictor $\boldsymbol{\varepsilon}_\theta$ now also depends on $\mathbf{c}$, $\boldsymbol{\varepsilon}_\theta = \boldsymbol{\varepsilon}_\theta(\mathbf{z}_t, t, \mathbf{c})$, and the corresponding conditional training objective becomes
\begin{equation}\label{eq:training_objective_cond}
    \mathcal{L}_{c}(\theta) = \mathbb{E}_{\mathbf{z}_0, \boldsymbol{\varepsilon}, t, \mathbf{c}} \left[ \lVert \boldsymbol{\varepsilon} - \boldsymbol{\varepsilon}_\theta (\mathbf{z}_t, t, \mathbf{c})\rVert_2^2 \right].
\end{equation}

Once trained, $\boldsymbol{\varepsilon}_\theta$ can be used to sample from $q(\mathbf{z}_0\mid\mathbf{c})$ \cite{zhan2024survey}. In practice, conditioning is implemented by injecting the information contained in $\mathbf{c}$ into the \gls{NN}, similarly to how the timestep $t$ is encoded. Several strategies can be employed for this purpose, including attention mechanisms, addition, and concatenation \cite{vaswani2017attention}.

\section{Database}\label{s:database}
The test case considered in this work comprises a set of \gls{CFD} simulations of the NASA-\gls{CRM} aircraft, a geometry characterized by a contemporary supercritical transonic wing and fuselage representative of a wide-body commercial transport aircraft. Further geometric specifications are detailed in \cite{vassberg2008crm}.

The database, originally derived by \citet{sabater2022ddbb} and already employed in other studies \cite[e.g.][]{hines2023graph, catalani2026towards}, contains \gls{Cp} distributions obtained by solving the \gls{RANS} equations coupled with the Spalart-Allmaras turbulence model \cite{allmaras2012sa} using the DLR TAU solver \cite{kroll2014tau}. The database includes variations in Mach number $M\in 
[0.502,0.897 ]$ and  Angle of Attack $\alpha \in[-2.47^\circ, 4.91^\circ] $, together with control surface deflections: Inboard
Aileron Deflection Angle $\delta_1 \in [-19.9^\circ, 19.4^\circ]$, Outboard Aileron Deflection Angle $\delta_2 \in [-19.95^\circ, 9.76^\circ]$, Elevator Deflection Angle $\delta_3 \in [-9.98^\circ, 9.83^\circ]$, and Horizontal Tailplane Deflection Angle $\delta_4 \in [-1.98^\circ, 1.93^\circ]$, resulting in $b=6$ governing parameters.

\begin{figure}[htb]
    \centering
    \includegraphics[width=8.2cm]{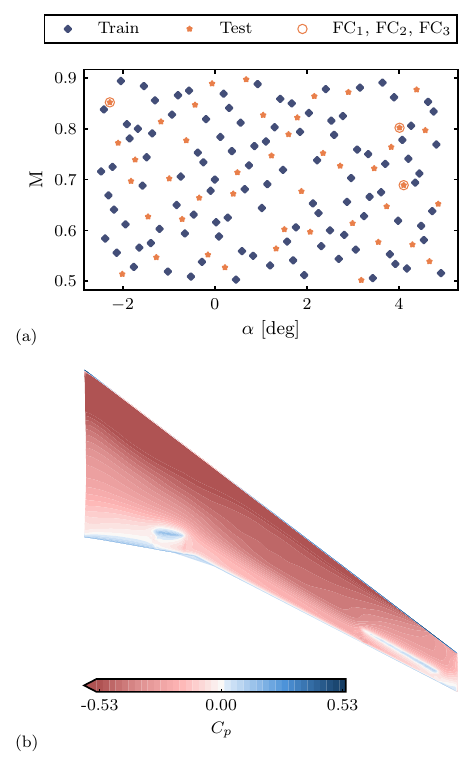}
    \caption{(a) Flight envelope for the NASA-\gls{CRM} configuration in the \gls{M}-\gls{AoA} plane. (b) Representative \gls{Cp} distribution over the upper surface of the wing, corresponding to the flight condition vector (\gls{M}, \gls{AoA}, \gls{Inail},\gls{Outail}, \gls{Elev}, \gls{Htp})$=(0.6, 2.5^\circ, -8.6^\circ, -8.0^\circ, -6.36^\circ, -1.38^\circ)$.}
    \label{fig:ma_envelope}
\end{figure}

A total of $n = 149$ parameter combinations were computed. Following the methodology proposed in \cite{sabater2022ddbb}, the dataset is partitioned such that 70\% is allocated for model \textit{training} (merging the original training and validation sets), while the remaining 30\% is reserved for \textit{testing}. The \gls{M}-\gls{AoA} envelope for both sets is illustrated in \autoref{fig:ma_envelope}a. The remaining parameter combinations are omitted for clarity but can be found in \cite{sabater2022ddbb}. Three flight conditions are selected from the test set to visualize the predictions over the wing, covering both low-transonic and high-transonic flow regimes. These flight conditions, whose parameter combinations can be found in \autoref{tab:visualization_test_cases}, are referred to as FC$_1$, FC$_2$, and FC$_3$, and are identified by circles within the \gls{M}-\gls{AoA} envelope in \autoref{fig:ma_envelope}a. 

\begin{table*}[htbp]
\centering
\caption{Sextuplet vector of the selected visualization test cases. All angles are expressed in degrees.}
\label{tab:visualization_test_cases}
\begin{tblr}{
  colspec = {QS[table-format=1.3]S[table-format=+1.3]S[table-format=+2.3]S[table-format=+2.3]S[table-format=+1.3]S[table-format=+1.3]},
  row{1} = {c},
  cell{2}{1} = {c},
  cell{3}{1} = {c},
  cell{4}{1} = {c},
  hline{1,5} = {-}{0.1em},
}
\textbf{Flight condition} & \textbf{M} & $\boldsymbol{\alpha}$ & $\boldsymbol{\delta_1}$     & $\boldsymbol{\delta_2}$      & $\boldsymbol{\delta_3}$     & $\boldsymbol{\delta_4}$     \\
FC$_1$                         & 0.689   & 4.100     & 4.000  & -0.510  & -2.220 & -1.100 \\
FC$_2$                         & 0.802   & 4.010     & 12.100 & -12.750 & 8.200  & -1.460 \\
FC$_3$                         & 0.852   & -2.280    & -4.300 & -17.550 & -4.370 & -0.220 
\end{tblr}
\end{table*}

Because the wing represents the most critical region for \gls{Cp} prediction compared to other aircraft elements, only the wing surface mesh is retained for model training, as shown in \autoref{fig:ma_envelope}b. This restriction enables the model to prioritize the capture of high-gradient features, such as the leading edge suction peak and shock waves, which are particularly challenging to resolve due to their strong nonlinearity. This discretization results in a point cloud of $h=139{,}374$ mesh points forming an unstructured surface representation. Consequently, the data matrix is defined as $\mathbf{D} \in \mathbb{R}^{n \times h}$, where each row corresponds to a single \gls{Cp} sample associated with a specific flight condition defined by the sextuplet (\gls{M}, \gls{AoA}, \gls{Inail}, \gls{Outail}, \gls{Elev}, \gls{Htp}). Thus, \gls{Cp}$_i \in \mathbb{R}^h$, with $i = 1, 2, \ldots, \ n$. In the following, \gls{Cp}$_i$ samples and their associated sextuplet information vectors are denoted by the symbols $\mathbf{x} \in \mathbb{R}^h$, and $\mathbf{c} \in \mathbb{R}^b$.

\section{Methodology}\label{s:methodology}
The proposed surrogate framework follows a three-stage procedure to predict \gls{Cp} distributions over the wing surface. First, \gls{PCA} is used as a pre-processing stage that reduces the surface pressure in a compact modal space. Second, a conditional \gls{DDPM} is trained on the resulting modal coefficients, learning to generate valid latent vectors from Gaussian noise. Finally, an inverse \gls{PCA} transformation maps the generated coefficients back to the physical domain. Each stage is detailed below.

\subsection{Data representation and dimensionality reduction}
%
To render diffusion-based modeling tractable for high-fidelity aerodynamic data, the surface pressure fields are first represented in a modal coordinate system using \gls{PCA}. This pre-processing step projects the high-dimensional wing pressure distributions onto an orthogonal basis, which improves training stability and significantly reduces computational cost \cite{tran2024electric, frances2024betavae}. Consider a dataset $\mathbf{X} = [\mathbf{x}_1, \dots, \mathbf{x}_d]^\top \in \mathbb{R}^{d \times h}$, where $d$ denotes the number of samples (either training, testing, or an arbitrary number), and $h$ is the number of mesh points on the wing surface. 
For each physical sample $\mathbf{x}_i \in \mathbb{R}^h$, the modal representation is denoted by $\mathbf{z}_i = \mathcal{F}(\mathbf{x}_i)$, where $\mathbf{z}_i \in \mathbb{R}^m$ and $m$ is the number of retained components.

In this work, \gls{PCA} is not used as a heavily truncated compression technique, but as a full, non-truncated, reversible linear reparameterization of the dataset representation. Retaining all components ensures that no information is discarded prior to the generative process, so that any reduction in reconstruction fidelity is attributable solely to the diffusion model rather than to modal truncation. The mapping between physical space and modal space is therefore exactly reversible for the dataset representation employed in this work. Since the maximum rank of the dataset is bounded by the number of training samples, $m = 105$ components are retained, corresponding to the size of the training set. This transformation reduces the representation from the full surface grid ($h$ points) to the intrinsic dimensionality supported by the dataset without loss of information within that representation. Consequently, the generative diffusion process is conducted entirely within this compact latent space, producing the latent matrix $\mathbf{Z} \in \mathbb{R}^{d \times m}$. The physical pressure distributions are subsequently recovered through the inverse transformation $\mathbf{x}_i = \mathcal{F}^{-1}(\mathbf{z}_i)$.

This modal-space formulation is adopted as a practical representation strategy for diffusion-based surrogate modeling on unstructured aerodynamic data. By operating on modal coefficients rather than rasterized fields, it avoids the interpolation and topology distortions that may arise when irregular surface data are mapped onto structured grids. Furthermore, compared to graph-based diffusion formulations, it provides a significantly more computationally tractable architecture that is well-suited for limited-data settings. Ultimately, this linear coordinate transformation allows a standard fully-connected diffusion model to learn the underlying physics without relying on complex, geometry-specific network topologies

\subsection{Network architecture}\label{ss:architecture}
The network architecture employed in this study is based on the U-Net, originally proposed by \citet{ronneberger2015unet} for biomedical image segmentation. While the standard U-Net relies on convolutional \glspl{NN} with spatial \textit{max-pooling}, \textit{stride}, and \textit{same} convolutions to extract hierarchical features from structured grids, the proposed generative process operates entirely on modal coefficients, making conventional spatial convolutions inapplicable. While graph-based architectures have been successfully applied to the same unstructured mesh \cite{hines2023graph}, their higher computational burden would compound the already significant cost of the iterative diffusion sampling process \cite{li2025research}. Therefore, to balance generative capacity with computational tractability, the contracting and expanding paths are constructed entirely with \textit{fully-connected} layers, leveraging conventional \glspl{MLP} to process the modal coefficients.

\subsubsection{Feature projection and backbone}
\begin{figure*}[ht]
    \centering
    \def\svgwidth{0.9\linewidth}
    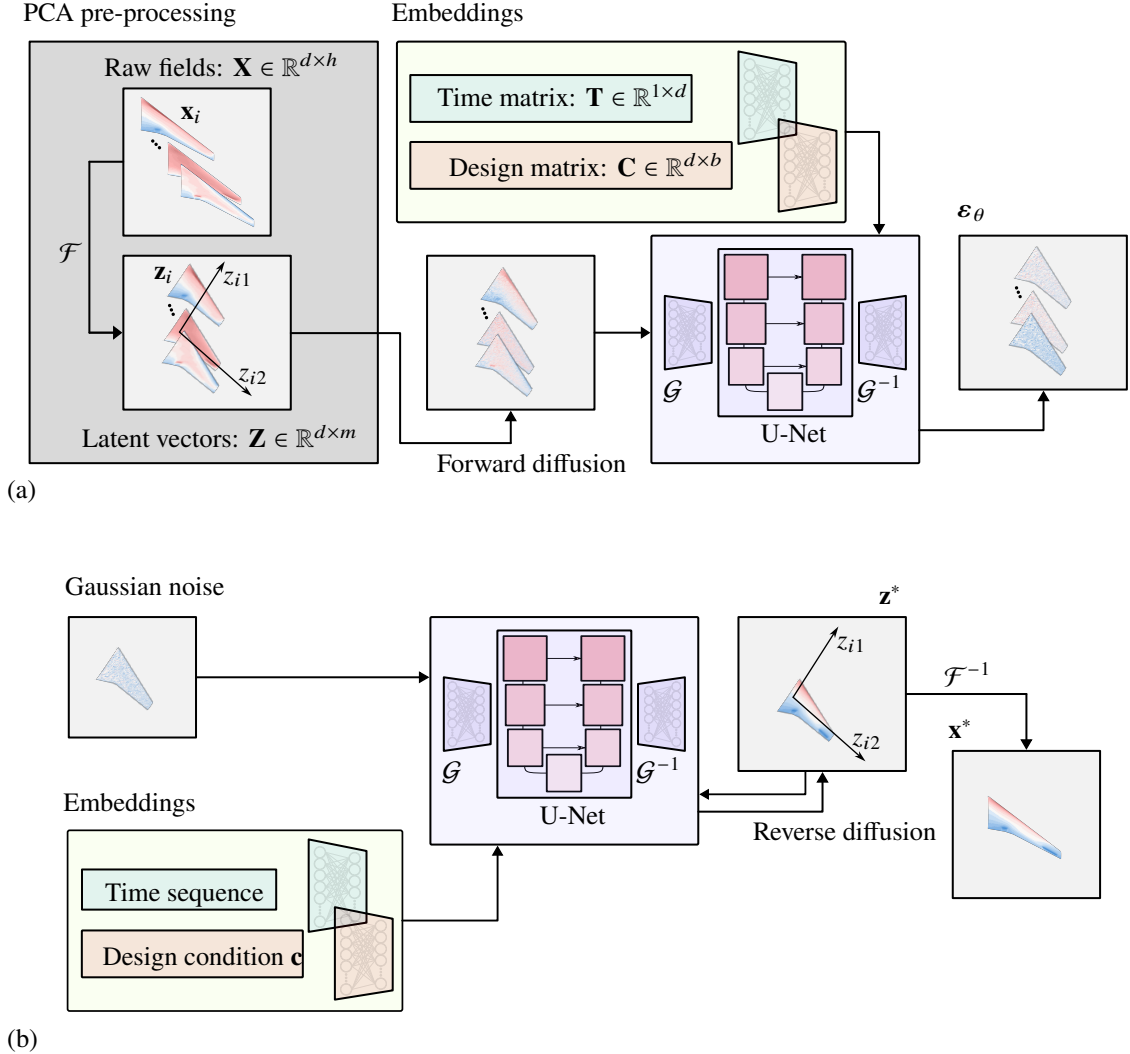
    \caption{(a) Training process of the surrogate. (b) Conditional sample generation from Gaussian noise.}
    \label{fig:architecture}
\end{figure*}

The training and inference processes are illustrated in \autoref{fig:architecture}. The input latent vector $\mathbf{z}_i$ is noised at a random noise level $t$, and then processed by a learnable linear projection layer $\mathcal{G}: \mathbb{R}^m \rightarrow \mathbb{R}^r$ that maps it to a fixed-dimensionality feature space of dimension $r$, which facilitates consistent processing by the network. The backbone comprises three hierarchical stages:
\begin{itemize}
    \item \textbf{Encoder (compression path):} progressively reduces the width of the feature vectors, compressing the input into a compact contextual representation. It is formed by \textit{down-sampling} blocks.
    \item \textbf{Bottleneck:} processes the data in its most compressed state, capturing global dependencies. It is formed by a \textit{middle} block.
    \item \textbf{Decoder (reconstruction path):} restores the feature dimensionality, concatenating processed features with stored states from the encoder through skip connections to recover fine-grained details lost during compression. It is formed by \textit{up-sampling} blocks.
\end{itemize}

Internally, the three classes of blocks are built from smaller modules with specific roles. These include \textit{residual} blocks, composed of fully-connected layers, Layer Normalization \cite{ba2016layernorm}, and ELU activations. In addition, self-attention blocks are employed to implement the self-attention mechanism \cite{vaswani2017attention}, enabling the modeling of complex interdependencies among the \gls{PCA} coefficients. The U-Net output then undergoes an inverse projection $\mathcal{G}^{-1}$ to yield the predicted noise $\boldsymbol{\varepsilon}_\theta \in \mathbb{R}^{m}$.

During inference, as illustrated in \autoref{fig:architecture}b, the model leverages the reverse diffusion process to generate new samples from pure Gaussian noise. The resulting vector $\mathbf{z}^*$, where the superscript $*$ denotes a model prediction, is then mapped back to the physical domain through $\mathcal{F}^{-1}$ to recover the pressure distribution $\mathbf{x}^*$.

Table \ref{tab:ddpm_characteristics} summarizes the architectural specifications and training hyperparameters. A linear noise schedule was selected following the variance progression proposed by \citet{ho2020ddpm}, as it provided the most stable behavior in preliminary experiments relative to cosine-based schedules and wider variance ranges \cite{nichol2021improved, guo2025noise}. Finally, training stability was further supported through weight decay \cite{kingma2014adam} and scheduled \gls{LR} \cite{smith2017super}.

\begin{table}[t]
    \centering
    \caption{Hyperparameters and architectural specifications of the proposed \gls{DDPM} framework.}
    \label{tab:ddpm_characteristics}
    \begin{tblr}{
      hline{1,18} = {-}{0.1em},
    }
    \textbf{Architecture} & \\
    {\,\,\,\,\,\,Activation} & ELU \\
    {\,\,\,\,\,\,Projection size ($r$)} & 1024 \\
    {\,\,\,\,\,\,Network depth} & 5 \\
    {\,\,\,\,\,\,Feature width sequence} & 512-256-128-64-32 \\
    \textbf{Diffusion parameters} & \\
    {\,\,\,\,\,\,Diffusion steps ($T$)} & 1000 \\
    {\,\,\,\,\,\,Noise schedule} & Linear \\
    {\,\,\,\,\,\,$\beta_{\text{start}}$} & \num{1e-04} \\
    {\,\,\,\,\,\,$\beta_{\text{end}}$} & \num{2e-02}\\
    \textbf{Training configuration} & \\
    {\,\,\,\,\,\,Optimizer} & Adam \\
    {\,\,\,\,\,\,Weight decay} & \num{1.2e-04} \\ 
    {\,\,\,\,\,\,\gls{LR} schedule} & OneCycle \\
    {\,\,\,\,\,\,\gls{LR}$_{\text{start}}$} & \num{1e-04} \\
    {\,\,\,\,\,\,\gls{LR}$_{\text{end}}$} & \num{1e-06} \\ 
    \textbf{No. of parameters} & 32,279,433
    \end{tblr}
\end{table}

\subsubsection{Conditioning embeddings}
The denoising network must infer the noise component not only from the corrupted latent state, but also from the context in which that state is observed, as outlined in \autoref{ss:conditioning}. In particular, the reverse process depends on two complementary sources of conditioning information: the diffusion timestep $t$, which specifies the current noise level, and the flight condition vector $\mathbf{c}$, which specifies the aerodynamic operating point associated with the target pressure distribution. To inject both sources of information into the backbone in a compatible form with its internal feature space, two embedding strategies are employed.

On the one hand, the \textbf{time embedding} encodes the temporal state of the diffusion process through a sinusoidal embedding \cite{vaswani2017attention}, which maps each discrete timestep $t$ into a high-dimensional vector of $k$ frequencies, allowing the network to identify the corresponding noise level. The time embedding is computed from the time matrix $\mathbf{T} \in \mathbb{R}^{1 \times d}$. In this work, the time embedding dimension is set to $k = 128$.

On the other hand, the \textbf{label embedding} provides a representation of the flight conditions, given by the sextuplet $\mathbf{c}$. The sextuplets are gathered in the design matrix $\mathbf{C} \in \mathbb{R}^{d \times b}$. The same embedding technique used for the time information is applied to each of the six design parameters, although the dimensionality is allocated hierarchically: $k = 64$ for the primary drivers (\gls{M}, \gls{AoA}), and $k = 32$ for the control surfaces (\gls{Inail}, \gls{Outail}, \gls{Elev}, \gls{Htp}).

Each embedding is passed through a learnable projection, improving the integration of temporal and physical information with the signal features propagated through the network. These embeddings are then injected into all down-sampling, middle, and up-sampling blocks. Ultimately, each input sample of principal components is associated with a 256-dimensional label embedding vector and a 128-dimensional time embedding vector, which jointly condition the diffusion process.

\subsection{Signal-aware training objective}\label{ss:training_strategy}
As reviewed in \autoref{s:theoretical_background}, the standard conditional \gls{DDPM} formulation minimizes the discrepancy between the actual noise $\boldsymbol{\varepsilon}$ and the predicted noise $\boldsymbol{\varepsilon}_\theta$ through \autoref{eq:training_objective_cond}. While this objective has proven effective in image generation, it treats all timesteps and all spatial regions equally, which is ill-suited to aerodynamic pressure data where sharp gradient features such as shock waves occupy a small fraction of the domain but carry disproportionate physical importance. In particular, the standard noise-matching formulation was observed to produce high-frequency artifacts, especially in regions characterized by strong pressure gradients such as shock waves, thereby degrading the generalization capability of the model.

To place greater emphasis on the reconstruction of the underlying physical signal, a signal-aware objective is introduced by measuring the discrepancy in signal space after applying the forward noising relation. Specifically, instead of directly penalizing the error between $\boldsymbol{\varepsilon}$ and $\boldsymbol{\varepsilon}_\theta$, the following objective is proposed:

\begin{equation}\label{eq:training_objective_mod}
    \mathcal{L}_{c}'(\theta) = \mathbb{E}_{\mathbf{z}_0, \boldsymbol{\varepsilon}, t, \mathbf{c}} \left[ \lVert \mathbf{z}_t(\boldsymbol{\varepsilon}) - \mathbf{z}_t(\boldsymbol{\varepsilon}_\theta) \rVert_2^2 \right],
\end{equation}
where $\mathbf{z}_t(\boldsymbol{\varepsilon})$ denotes the noisy signal obtained at timestep $t$ using the actual noise, and $\mathbf{z}_t(\boldsymbol{\varepsilon}_\theta)$ denotes the corresponding noisy signal obtained when the noise predicted by the U-Net is used instead. In this way, the optimization is formulated in terms of the discrepancy induced on the corrupted signal, rather than solely in terms of the raw noise mismatch. Thus, this formulation ensures that the model captures the essential topological and physical properties of the pressure distributions, rather than merely fitting the statistical properties of the noise.

Because \autoref{eq:forward_process_alphabar} is used to generate $\mathbf{z}_t$ in both cases, and the timestep $t$ and clean sample $\mathbf{z}_0$ are common for both expressions, the proposed objective can be simplified as
\begin{equation}\label{eq:training_objective_mod_simp}
    \mathcal{L}_{c}'(\theta) =  \mathbb{E}_{\mathbf{z}_0, \boldsymbol{\varepsilon}, t, \mathbf{c}} \left[ \lVert \sqrt{(1 - \overline{\gamma}_t)} \cdot [\boldsymbol{\varepsilon} - \boldsymbol{\varepsilon}_\theta (\mathbf{z}_t, t, \mathbf{c})]\rVert_2^2 \right].
\end{equation}

The proposed formulation is therefore equivalent to a timestep-dependent weighted variant of the standard loss in \autoref{eq:training_objective_cond}, where the weighting factor $\sqrt{1-\overline{\gamma}_t}$ modulates the contribution of each timestep according to its influence on the corrupted signal. This factor increases monotonically with t, assigning greater emphasis to the heavily noised timesteps where large-scale signal structure must be recovered. In the final diffusion steps, $t \to T$, where the signal is heavily degraded, the weighting term approaches unity, encouraging the network to better recover the large-scale structure of the noised pressure field. Conversely, as the process approaches the clean signal, $t \to 0$, the weighting term decreases, reducing the relative influence of the final denoising steps. In practice, this redistribution of emphasis was found to mitigate high-frequency artifacts and to produce smoother and more physically consistent pressure distributions. 

\section{Results}\label{s:results}
The predictive performance of the proposed surrogate framework is evaluated in this section. To ensure a robust comparative analysis, the assessment benchmarks the proposed method against three alternative architectures. To maintain consistency and ensure a fair comparison, all evaluated models employ the same \gls{PCA} pre-processing stage, thereby operating within an identical modal subspace. Consequently, the ``PCA+'' prefix is omitted from the model nomenclature for brevity. The evaluated architectures are categorized as follows: 

\begin{itemize}
    \item \textbf{External} baselines: two established architectures serve as reference points. The first, denoted as \textit{MLP}, consists of a standard \gls{MLP} that serves as a direct regression benchmark, mapping the input flight-condition sextuplet directly to the principal components of the pressure field. The second baseline, referred to as \textit{AE+GPR}, adopts an \gls{AE}-based methodology. This approach incorporates a \gls{GPR} to model the mapping between flight parameters and the latent space, a strategy whose efficacy has been demonstrated in recent aerodynamic studies \cite{frances2024betavae, solera2024betavae, tran2024electric}.

    \item \textbf{Internal} baseline and proposed method: to isolate the specific impact of the proposed training objective (see \autoref{ss:training_strategy}), the standard formulation of the diffusion model is included as an internal baseline. This variant is referred to as \textit{DDPM-N}, whereas the proposed model, which follows the signal-aware formulation, is denoted as \textit{DDPM-S}.
\end{itemize}

The comparison is scoped to isolate the value of the proposed fully-connected diffusion formulation and, in particular, the effect of the signal-aware training objective. Architectures operating outside the same \gls{PCA}-space representation, such as graph-based, operator-learning, or neural-field approaches, are therefore not included, as their comparison would conflate the contribution of the representation with that of the generative framework.

While the \textit{DDPM-N} variant mirrors the architecture of the proposed \textit{DDPM-S}, the architectural depth and complexity of the external baselines are specifically constrained to account for the limited training data available ($m=105$). Both the \textit{MLP} and the \textit{AE+GPR} frameworks are constructed using shallow, fully-connected layers to reduce susceptibility to memorization in this sparse-data regime, thereby providing a stable and conservative benchmark. As identified in studies of limited-sample learning, overly complex architectures are highly susceptible to overfitting when the number of trainable parameters exceeds the information content of the training samples \cite{hawkins2003overfitting, ying2019overfitting}. 

The analysis is structured as follows: first, a comparative assessment is performed across all four surrogate models. Second, a focused characterization of the \textit{DDPM-S} is conducted, comprising a \gls{CH} integrity check with respect to the design parameters, and a sensitivity analysis involving the variance of the model. 

\subsection{Error distribution}\label{ss:error_distribution}
\begin{figure}[t]
    \centering
    \includegraphics[width=8.2cm]{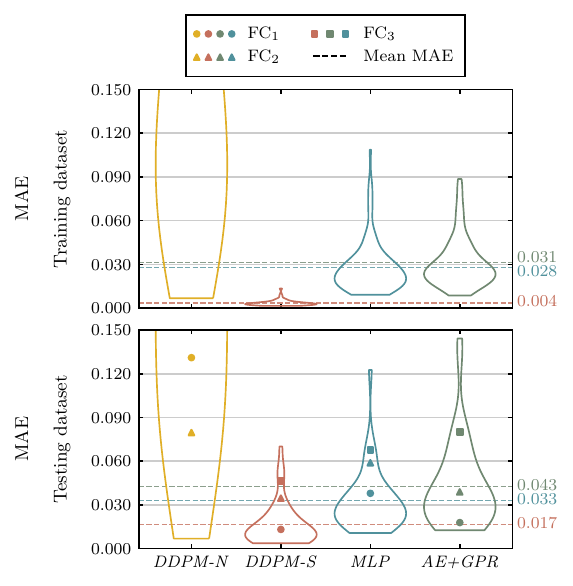}
    \caption{\gls{MAE} distribution between actual and predicted \gls{Cp} fields for the different models and database sets. Visualization test cases are denoted by symbols. Dashed lines represent the global mean \gls{MAE} for each model. Note that the distribution tails are truncated at the sample extrema (minimum and maximum observed errors).}
    \label{fig:violin_plot}
\end{figure}

A quantitative assessment of the surrogate frameworks is conducted using the \gls{MAE} metric, which evaluates the deviation between the ground-truth \gls{Cp} distributions and model predictions across both training and testing datasets. \autoref{fig:violin_plot} illustrates the comparative error distributions, with specific emphasis on the selected visualization test cases FC$_1$, FC$_2$, and FC$_3$ (see \autoref{tab:visualization_test_cases}).

Across both training and testing datasets, the proposed \textit{DDPM-S} achieves the lowest reconstruction errors among all evaluated architectures. Most notably, the standard \textit{DDPM-N} is substantially outperformed by \textit{DDPM-S}, this result suggests that the standard noise-prediction objective is less effective at capturing the physical dependencies encoded within the modal coefficients. The degradation of \textit{DDPM-N} is particularly severe for FC$_3$, the highest-complexity flight condition, where the model fails to reproduce the dominant pressure topology as discussed in subsection \ref{s:PFV}.

Regarding the external baselines, both exhibit competitive performance, but \textit{DDPM-S} attains superior global accuracy. This advantage is quantified through the mean value of the error distribution, as the proposed model yields a mean \gls{MAE} approximately 48\% lower than the \textit{MLP} and 60\% lower than the \textit{AE+GPR} architecture. This trend is also consistent across the three highlighted flight conditions, for which \textit{DDPM-S} exhibits the lowest relative error. Furthermore, the error distribution of the proposed model is structurally more favorable, being characterized by attenuated upper tails, which indicates fewer extreme outliers, and by a higher concentration of probability density within the low-error regime.

\subsection{Pressure field visualization} \label{s:PFV}
\begin{figure*}[htb]
    \centering
    \includegraphics[width=0.9\linewidth]{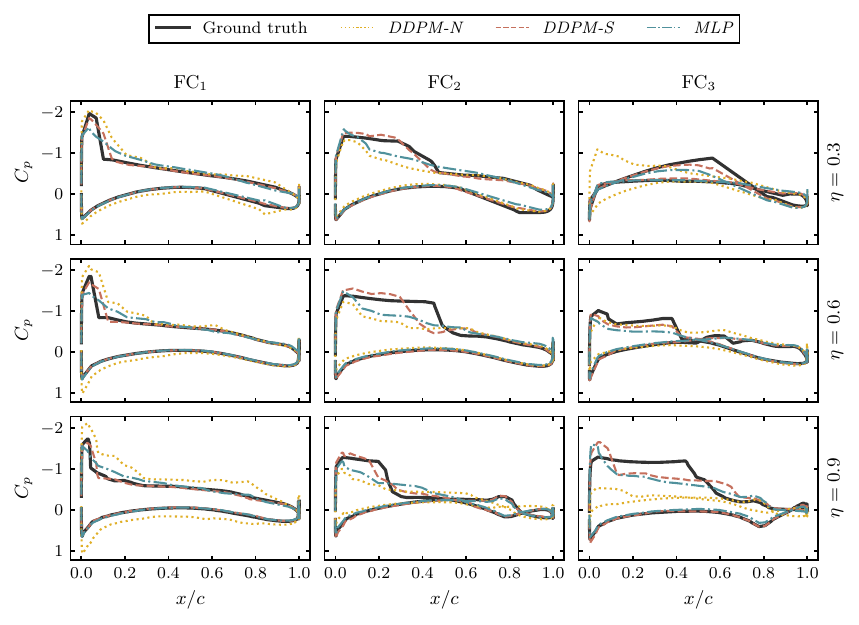}
    \caption{Chordwise pressure distributions for actual and surrogate \gls{Cp} fields across span percentages $\eta = 0.3, 0.6, 0.9$. The position over the airfoil is normalized with the chord $c$.}
    \label{fig:chordwise_distributions}
\end{figure*}

The predicted pressure distributions are examined for the three visualization test cases, retaining \textit{MLP} as the sole external baseline given its superior performance over \textit{AE+GPR}, while including both \textit{DDPM-S} and \textit{DDPM-N} to assess the effect of the signal-aware objective introduced in \autoref{ss:training_strategy}. The chordwise pressure distributions are shown in \autoref{fig:chordwise_distributions} for three normalized spanwise positions denoted by $\eta$, where $\eta=0$ identifies the wing root and $\eta=1$ the tip.

The lower wing surface, characterized by smoother pressure gradients, is generally well resolved by both \textit{DDPM-S} and \textit{MLP}, although control surface deflections introduce local discontinuities at specific spanwise locations, such as $\eta = 0.9$ for conditions FC$_2$ and FC$_3$. 

The upper surface provides a more discriminating comparison because it contains the strongest nonlinear flow features. In general, \textit{DDPM-S} shows improved fidelity in the reconstruction of these structures, particularly at the $\eta=0.6$ spanwise station across all three flight conditions. This behavior is especially evident in the leading edge suction peak, for which the \textit{MLP} baseline tends to over-smooth the pressure gradient and under-predict the peak magnitude relative to the ground truth, the \textit{DDPM-S} preserves the sharper profile. Similar differences are visible in the shock-related pressure jumps and plateau regions, where the proposed model more closely follows the reference distributions, particularly in FC$_2$ and FC$_3$.

In contrast, the \textit{DDPM-N} architecture fails to robustly reconstruct the global pressure topology in the most demanding cases. Although it yields reasonable approximations for the lower surface in some conditions and spanwise stations, such as $\eta=0.3$ and $0.6$ for FC$_1$ and FC$_2$,  its behavior deteriorates markedly as the flow complexity increases. For instance, in FC$_1$ at $\eta=0.9$, the \textit{DDPM-N} exhibits a strong bias and overestimates the suction peak. More critically, in the high-complexity case of FC$_3$, the model is unable to reproduce the dominant upper surface trends at the $\eta=0.3$ and $0.9$. This behavior is consistent with the beneficial effect of the signal-aware weighting in \autoref{eq:training_objective_mod_simp}, which redistributes the training emphasis toward timesteps that more strongly affect the reconstructed signal.

\begin{figure*}[t]
    \centering
    \includegraphics[width=0.9\linewidth]{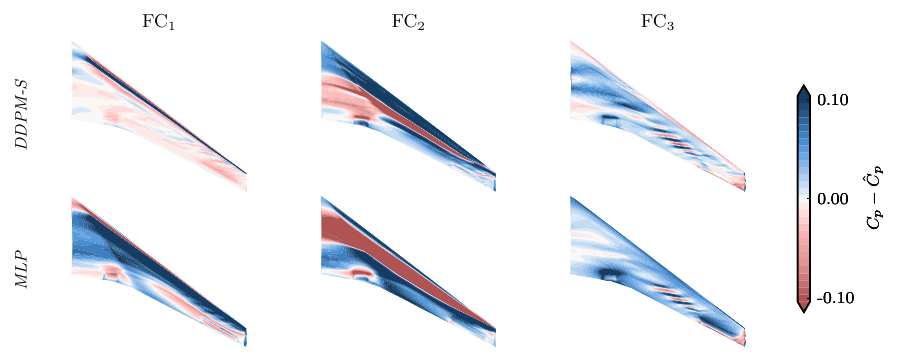}
    \caption{Error distribution over the upper surface of the wing for \textit{DDPM-S} (top row), and \textit{MLP} (bottom row) models.}
    \label{fig:contours_wing}
\end{figure*}

These qualitative observations are further supported by the upper surface error contours depicted in \autoref{fig:contours_wing}. To maintain a focused comparison between the most competitive models, the \textit{DDPM-N} is omitted from this visualization. For both architectures, the largest deviations are concentrated in regions with strong pressure gradients, particularly near shock waves and control surface hinge lines. This is most evident for FC$_2$ and FC$_3$, where the discontinuity at the inboard aileron boundary is clearly noticeable. However, the \textit{DDPM-S} exhibits systematically lower error levels in these critical regions, while also reducing the background error over the rest of the upper surface.

Overall, the visual evidence indicates that \textit{DDPM-S} more faithfully reconstructs the dominant nonlinear structures of the pressure field than the compared baselines, while maintaining a consistent improvement in field-level fidelity across the selected test conditions.

\subsection{Conditioning-space error distribution}
To examine how the surrogate error distributes across the multi-dimensional design space, the influence of the conditioning sextuplet (\gls{M}, \gls{AoA}, \gls{Inail}, \gls{Outail}, \gls{Elev}, \gls{Htp}) is analyzed. Because direct visualization of the full six-dimensional space is not possible, the analysis is decomposed into bi-variate projections, focusing on the \gls{M}-\gls{AoA}, \gls{Inail}-\gls{Outail}, and \gls{Elev}-\gls{Htp} sub-spaces. For each projection, the \glspl{CH} of the training and testing sets are delineated to visualize the domain boundaries. Superimposed on these projections are the three selected visualization test cases, together with the three testing samples yielding the highest and lowest reconstruction errors. Given the limited number of available flight conditions, this analysis is intended as a qualitative visualization of the error topology in conditioning space rather than as a formal density-based generalization study.

\begin{figure}[ht]
    \centering
    \includegraphics[width=8cm]{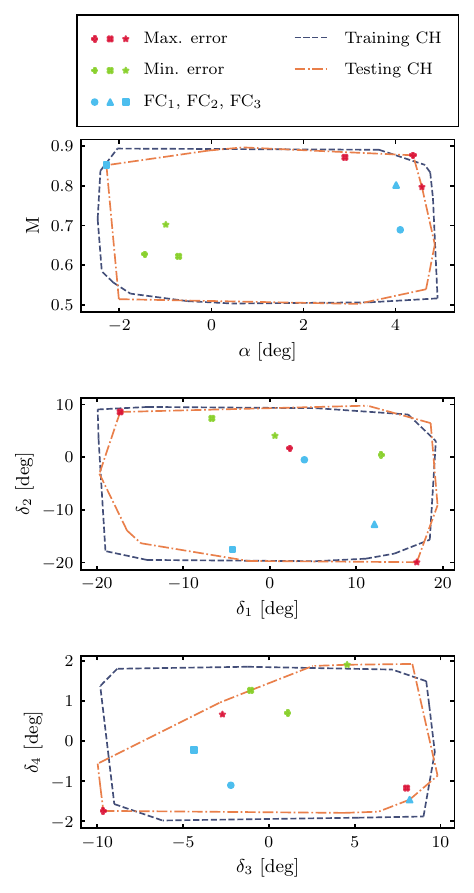}
    \caption{Visualization test cases together with the three maximum- and minimum-error flight conditions for the \textit{DDPM-S} within the training and testing \gls{CH} spaces.}
    \label{fig:fc_analysis}
\end{figure}

The spatial distribution of these samples, shown in \autoref{fig:fc_analysis}, reveals that the macroscopic complexity of the pressure field, and hence the surrogate error, is governed primarily by \gls{M} and \gls{AoA}. The cases with the highest reconstruction errors (marked in red) cluster along the maximum Mach number boundary of the \gls{M}-\gls{AoA} envelope, a region associated with transonic conditions in which shock waves and incipient separation become more prominent. The location of the visualization test cases is consistent with this trend, since FC$_3$, previously identified as a highly complex prediction case, lies directly on this upper \gls{M}-\gls{AoA} boundary. Conversely, the samples with the lowest errors (marked in green), as well as the other visualization cases, are located in the interior or near the lower boundary of the domain, corresponding to more benign aerodynamic conditions. 

In contrast, this strong spatial organization is not observed in the control surface projections. In the \gls{Inail}-\gls{Outail} plane, both high- and low-error samples are distributed without a clear pattern over the admissible range of deflections. The same behavior is even more evident in the \gls{Elev}-\gls{Htp} plane, where extreme error cases appear scattered along the \gls{CH} perimeter. These observations suggest that, although wing control surfaces induce localized flow modifications, they play a secondary role relative to the global flight conditions in determining the overall reconstruction difficulty, while empennage controls have only a limited impact on the wing pressure field. This behavior is also consistent with the architectural embedding strategy established in \autoref{ss:architecture}: a larger embedding dimension was assigned to \gls{M} and \gls{AoA}, allowing the model to place greater representational emphasis on the primary aerodynamic drivers without over-weighting secondary inputs.

\subsection{Sensitivity analysis}
The probabilistic formulation of the \gls{DDPM} is the primary driver behind the model's superior generalization capabilities, and its intrinsic stochasticity provides an additional degree of information that deterministic surrogates cannot offer. This approach inherently introduces stochasticity into the inference process, implying that a fixed conditioning input vector does not yield a deterministic output, but rather one realization of the learned conditional generative process, $p_\theta$. While alternative sampling schemes \cite{song2022ddim} are often employed to reduce stochastic variability and accelerate generation, exploiting this intrinsic variability is a central objective of present study. Accordingly, this subsection characterizes the sampling-induced spread of the model predictions in order to assess the stability of the generated solutions, and to investigate whether this spread provides useful qualitative information about regions and conditions associated with larger surrogate errors.

\subsubsection{Statistical convergence of \texorpdfstring{$p_\theta$}{ptheta}}
Establishing a practical inference strategy first requires determining the ensemble size $N$ needed to characterize the stochastic behavior of the \gls{DDPM} on deterministic pressure fields. While the previous analyses in this work relied on a single sample from $p_\theta$ (that is, $N = 1$), the study of sampling-induced variability requires the evaluation of larger ensembles. To this end, a pool of $N_{\text{max}} = 1000$ independent samples is generated for each of the $l = 44$ testing flight conditions.

For any evaluated ensemble size $N \in \{1, 2, \dots, 1000\}$, a random subset of $N$ predictions is sampled from this pool. Let $\mathbf{x}^*_{k,i} \in \mathbb{R}^h$ denote the $k$-th sampled prediction for the $i$-th flight condition, where $k = 1, \ldots, N$, and $i = 1, \ldots, l$. First, the element-wise ensemble mean $\boldsymbol{\mu}_{\theta, i}^{(N)} \in \mathbb{R}^h$ and standard deviation $\boldsymbol{\sigma}_{\theta, i}^{(N)} \in \mathbb{R}^h$ are computed across the $N$ samples for each specific flight condition:

\begin{align}
    \boldsymbol{\mu}_{\theta, i}^{(N)} & = \frac{1}{N} \sum_{k=1}^N \mathbf{x}^*_{k,i},\\
    \boldsymbol{\sigma}_{\theta, i}^{(N)} & = \sqrt{ \frac{1}{N} \sum_{k=1}^N \left( \mathbf{x}^*_{k,i} - \boldsymbol{\mu}_{\theta, i}^{(N)} \right)^2 }.
\end{align}

Next, these aggregated fields are collapsed over the $h$ spatial grid points to obtain scalar metrics per flight condition. The error is quantified by the \gls{RMSE} between the ensemble mean and the ground-truth pressure field $\mathbf{x}_i$, while the predictive spread is quantified by the spatial average of the standard deviation field. Let $j$ denote the spatial index of the mesh points:

\begin{align}
  \text{RMSE}_i^{(N)} & = \sqrt{ \frac{1}{h} \sum_{j=1}^h \left( x_{i,j} - \mu_{\theta, i, j}^{(N)} \right)^2 },\\
  s_i^{(N)} & = \frac{1}{h} \sum_{j=1}^h \sigma_{\theta, i, j}^{(N)}.
\end{align}

Finally, to yield the ultimate scalar values representing the entire test set for a given ensemble size $N$, the metrics are averaged across all $l$ flight conditions:

\begin{align}
    \overline{\text{RMSE}}^{(N)} & = \frac{1}{l} \sum_{i=1}^l \text{RMSE}_i^{(N)},\\
    \overline{\sigma}^{(N)} & = \frac{1}{l} \sum_{i=1}^l s_i^{(N)}.
\end{align}

By repeating this procedure across multiple values of $N$, a coupled relationship between the number of generative samples and the aggregated model performance is established.

\begin{figure}[t]
    \centering
    \includegraphics[width=8cm]{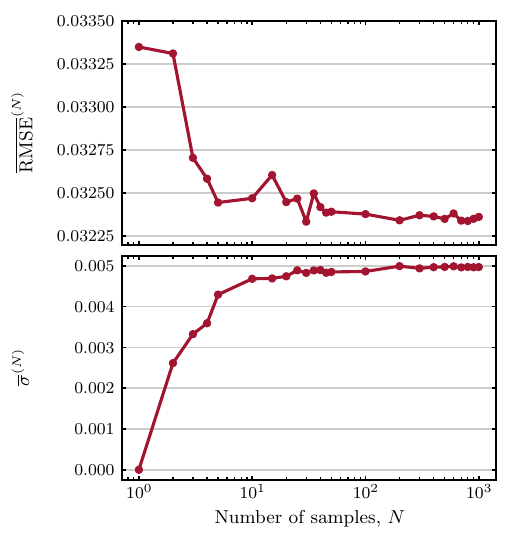}
    \caption{Convergence of the model's statistics as a function of ensemble size $N$. Bar symbol denotes mean values across the $l$ flight conditions.}
    \label{fig:rmse_std}
\end{figure}

\autoref{fig:rmse_std} illustrates the evolution of both metrics as a function of $N$. The error evolution (top panel) reaches a plateau of approximately $\overline{\text{RMSE}}^{(N)} \approx 0.0324$ for $N \geq 50$, corresponding to an accuracy improvement of roughly 3\% relative to the single-sample baseline. This interpretation is consistent with the convergence of $\overline{\sigma}^{(N)}$ (bottom panel), which stabilizes at approximately $0.005$ for $N \geq 50$. The low magnitude and rapid stabilization of both quantities indicate that the deterministic structure of the pressure fields is largely preserved, despite the intrinsic stochasticity of the \gls{DDPM}. Together, these results indicate that single-sample inference is a reasonable practical choice for point prediction, while ensembles of $N \geq 50$ samples are sufficient for reliable characterization of the sampling-induced spread.

The bounded spread observed in \autoref{fig:rmse_std} is consistent with the role of the proposed signal-aware training objective, which prioritizes the reconstruction of the underlying physical signal over the standard noise-matching loss. In practice, this behavior indicates that the modified formulation retains the beneficial stochasticity of the diffusion process while limiting the dispersion of the generated solutions around the deterministic aerodynamic trends learned from the data. In that sense, the sampling variability remains sufficiently structured to support the subsequent sensitivity analysis, without dominating the reconstructed pressure fields.

\subsubsection{Spatial deviation}
\begin{figure*}[t]
    \centering
    \includegraphics[width=0.9\linewidth]{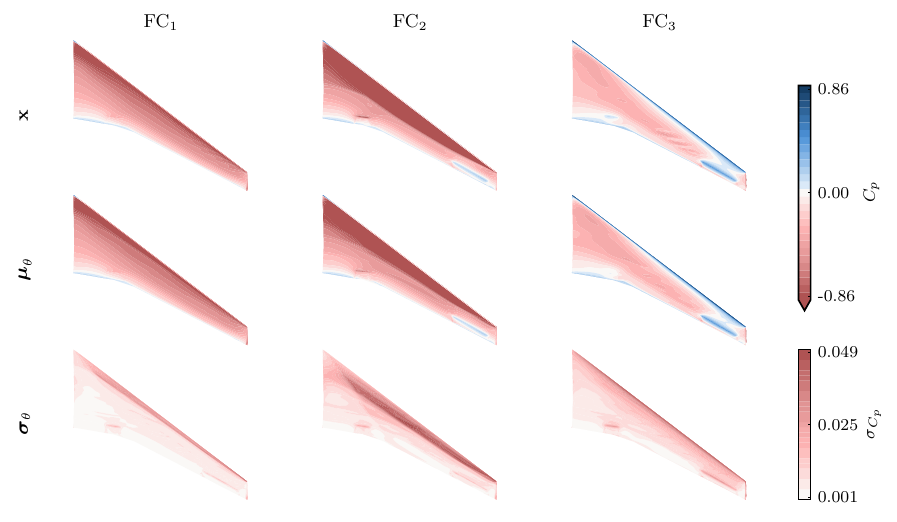}
    \caption{Spatial deviation of the model. The rows display the actual fields ($\mathbf{x}$) along with the ensemble means ($\boldsymbol{\mu}_\theta$) and standard deviations ($\boldsymbol{\sigma}_\theta$) computed from $N = 1000$ independent samples for each flight condition.}
    \label{fig:sensitivity_study}

\end{figure*}

For each of the three visualization test cases, $\boldsymbol{\mu}_\theta \in \mathbb{R}^h$ and $\boldsymbol{\sigma}_\theta \in \mathbb{R}^h$ are computed from the $N_{\text{max}} = 1000$ generated samples and plotted over the wing, providing a spatial view of the model's sampling-induced variability. \autoref{fig:sensitivity_study} compares the ground truth field $\mathbf{x}$, the ensemble mean $\boldsymbol{\mu}_\theta$, and the standard deviation contours $\boldsymbol{\sigma}_\theta$. For the three considered conditions, the ensemble mean remains very close to the reference field, indicating that the dominant deterministic structure of the pressure distribution is preserved after averaging over the stochastic samples. 

The standard deviation maps $\boldsymbol{\sigma}_\theta$ provide further insight into how the model's sampling spread is distributed over the wing. This spread  is not spatially uniform, but instead concentrates in regions characterized by strong flow gradients and geometric discontinuities:
\begin{itemize}
    \item \textbf{Control surfaces:} in all cases, elevated spread is observed along the hinge lines of the inboard and outboard ailerons. This indicates a greater sensitivity of the generated solutions in resolving the sharp pressure jumps associated with control surface deflections.
    \item \textbf{Leading edge:} the suction peak region exhibits noticeable spread. Given the extreme nonlinearity of the pressure gradient near the leading edge, the model shows higher sensitivity in reconstructing the exact peak magnitude.
    \item \textbf{Shock waves:} within the transonic regime, the spread forms high-intensity bands aligned with the shock fronts. Since the location and strength of a shock are highly sensitive to marginal changes in the input operating conditions, these regions naturally emerge as the dominant contributors to the spatial variability of the generated fields.
\end{itemize}

Consequently, the spatial distribution of $\boldsymbol{\sigma}_\theta$ can be interpreted as a qualitative map of modeling sensitivity. Regions associated with stronger flow complexity, such as shock structures, suction peaks, and hinge-line discontinuities, are also the regions where the generative process exhibits the largest spread. In this sense, the spatial deviation does not provide a calibrated uncertainty estimate, but it does offer useful reliability information for identifying where larger surrogate errors are more likely to occur and where high-fidelity validation may be most informative.

\subsubsection{Deviation-Error alignment}
Given that the predicted standard deviation tends to accumulate in aerodynamic regions characterized by strong nonlinearities, a deeper investigation into its spatial organization is conducted. Specifically, the localized alignment between the predicted standard deviation, $\boldsymbol{\sigma}_\theta$, and the absolute prediction error, $\mathbf{e} \in \mathbb{R}^h= |\boldsymbol{\mu}_\theta - \mathbf{x}|$, is analyzed.

For a given flight condition, let the set of $h$ spatial mesh points be sorted in ascending order of their standard deviation values. Let $S_\omega$ be a subset containing the fraction $\omega \in (0, 1]$ of these sorted mesh points, that is, the fraction of the associated wing with the lowest predicted spread. The dependence of $\mathbf{e}$ on $\boldsymbol{\sigma}_\theta$ is quantified through the \gls{LRI}, defined as the \gls{MAE} computed exclusively over the subset $S_\omega$:
\begin{equation}
    \text{LRI}(\omega) = \frac{1}{|S_\omega|} \sum_{i \in S_\omega} \mathbf{e}_i^{s},
\end{equation}
where $\mathbf{e}^s$ represents the error vector permuted to match the ascending order of $\boldsymbol{\sigma}_\theta$, and $|S_\omega| = \lfloor \omega \cdot h \rfloor$ is the cardinality of the subset. Conceptually, the \gls{LRI} can be interpreted as a dynamic \gls{MAE} evaluated over an expanding set of spatial points ranked by their predicted spread. As $\omega$ increases, regions with larger standard deviation are progressively incorporated into the error calculation. Consequently, evaluating the \gls{LRI} at the global limit ($\omega = 1$) exactly recovers the standard, full-field \gls{MAE} for the considered flight condition.

This formulation provides a diagnostic curve to assess the consistency between the model spread and the actual reconstruction error. Ideally, the \gls{LRI} should exhibit a monotonic increase with respect to $\omega$. Such behavior indicates that the model correctly assigns high confidence to accurate regions: the subset of points with the lowest spread, i.e. low $\omega$, corresponds to the lowest residuals. As the evaluation set $S_\omega$ expands to include points with higher predicted spread, the average error is expected to increase. Conversely, a flat or inverted curve would indicate that the model spread is only weakly related to the actual residual error.

\begin{figure}[t]
    \centering
    \includegraphics[width=8cm]{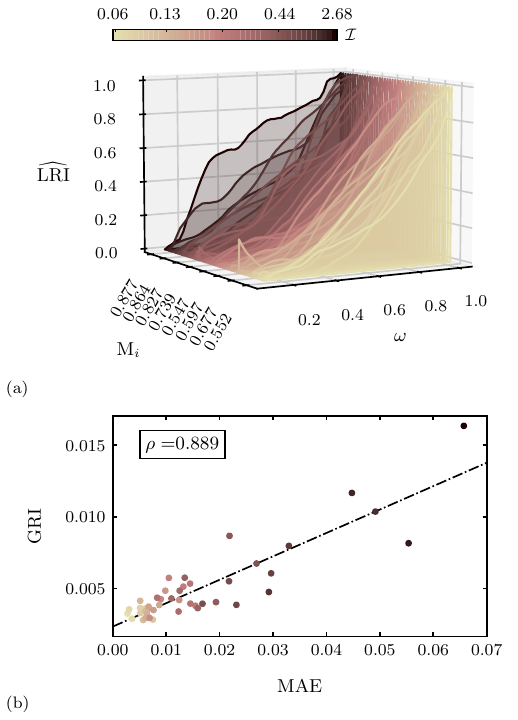}
    \caption{(a) LRI curves as a function of $\omega$ and the Mach number associated with the flight condition. (b) GRI-\gls{MAE}-$\mathcal{I}$ relationship.}
    \label{fig:lg_index}
\end{figure}

\autoref{fig:lg_index}a presents the $\omega$-\gls{LRI} profiles for the $l$ testing conditions. To facilitate the comparison, the \gls{LRI} is normalized to the unit interval $[0, 1]$ for each flight condition, denoted herein as $\widehat{\text{LRI}}$. In general, a consistent monotonic increase is observed across the dataset, confirming that regions of higher standard deviation also tend to exhibit higher  high reconstruction error. In addition, the $l$ curves are sorted and colored according to the scalar metric
\begin{equation}\label{eq:integral_lri}
    \mathcal{I} = \int_0^1 \text{LRI}(\omega) d\omega,
\end{equation}
which represents the area under the non-normalized curve. While the \gls{LRI} provides a spatially resolved view of the deviation-error relationship for each individual flight condition, the integral quantity $\mathcal{I}$ offers a global measure, effectively defining a compact scalar descriptor of the cumulative error distribution relative to the ranking induced by $\boldsymbol{\sigma}_\theta$. Notably, the magnitude of $\mathcal{I}$ exhibits a positive correlation with the main flight condition variables, particularly Mach number, as highlighted in the figure. This trend is consistent with the increasing difficulty of accurately resolving pressure fields as compressibility effects become more pronounced, thereby underscoring the heightened modeling complexity associated with high-transonic conditions.

The profile of the \gls{LRI} curves further elucidates the relationship between flow physics and surrogate sensitivity. For low-transonic regimes, most spatial mesh points exhibit both low spread and low reconstruction error, and the \gls{LRI} therefore increase gradually as $\omega$ approaches unity. In contrast, high-transonic regimes contain a larger proportion of points with elevated errors, primarily due to the presence of strong shock waves and localized compressibility effects. In these cases, the \gls{LRI} profile remains low for the high-confidence regions (small $\omega$) and steepens more abruptly as the high-spread regions are incorporated into the calculation. This divergence in the \gls{LRI} curve shapes indicates that the sampling-induced spread is not uniformly distributed, but instead concentrates around the most challenging aerodynamic features.

Additionally, the global relationship between $\mathbf{e}$ and $\boldsymbol{\sigma}_\theta$ is examined in \autoref{fig:lg_index}b, through the \gls{GRI}, a metric defined as the spatial mean of the element-wise standard deviation for each flight condition:
\begin{equation}\label{eq:gri}
    \text{GRI} = \frac{1}{h}\sum_j^h \sigma_{\theta, j}. 
\end{equation}

The \gls{GRI} therefore aggregates local spread information into a single scalar that can be evaluated at inference time without access to the ground truth. As illustrated in \autoref{fig:lg_index}b, this scalar exhibits a strong linear correlation with the reconstruction \gls{MAE} across the test set, with a Pearson coefficient of $\rho = 0.889$, indicating that the diffusion spread is a reliable qualitative indicator of prediction difficulty at the condition level. This dependence indicates that flight conditions with larger global sampling spread also tend to exhibit larger reconstruction errors in the present dataset.
Furthermore, the data points are colored according to the integral metric $\mathcal{I}$ (see \autoref{eq:integral_lri}), revealing a consistent trend: flight conditions with larger cumulative deviation-error misalignment (high $\mathcal{I}$) correspondingly exhibit elevated global spread (high \gls{GRI}) and reconstruction error. 

To verify that this self-awareness is a characteristic property of the diffusion process, rather than a trivial byproduct of model ensembling, a comparative experiment was conducted using the \textit{MLP} baseline. An ensemble of 200 independently initialized  \textit{MLP} models was used to generate a comparable spread metric. This approach yielded a markedly lower correlation coefficient of $\rho = 0.547$ for the \gls{GRI}-\gls{MAE} relationship. The disparity suggests that, whereas \textit{MLP} spread mainly reflects variability induced by weight initialization, the \gls{DDPM} spread encodes deeper physical information regarding the data distribution. This confirms that the generative framework not only produces superior predictions but also offers significantly more meaningful insights into predictive spread.

Ultimately, these results demonstrate that the proposed \textit{DDPM-S} transcends the limitations of traditional \textit{black-box} surrogates by providing a built-in, statistically grounded self-assessment mechanism. By leveraging its intrinsic stochasticity, the model offers a self-aware prediction capability to qualitatively assess where larger surrogate errors are more likely to occur. In this sense, the diffusion-sampling spread does not constitute calibrated uncertainty quantification. Still, it does act as a built-in sensitivity diagnostic that can help identify difficult spatial regions and challenging operating conditions in practical inference scenarios.

\section{Conclusion}\label{s:conclusion}
This work has presented a conditional DDPM surrogate for predicting three-dimensional aerodynamic surface pressure distributions over the NASA-\gls{CRM} wing in transonic conditions, demonstrating that diffusion-based generative models can deliver competitive field accuracy while providing built-in qualitative reliability information that deterministic surrogates cannot offer. The proposed \textit{DDPM-S} architecture is conditioned on a six-dimensional flight envelope comprising Mach number, angle of attack, and four control surface deflections, and operates entirely within a \gls{PCA}-based modal space that serves as a lossless reparameterization of the unstructured surface mesh.

By treating PCA as a full, lossless reparameterization rather than a truncated reduced-order model, the framework accommodates the unstructured surface mesh through fully-connected layers, avoiding the complexity of graph-based operators while preserving the complete information content of the training data. The training process incorporates a signal-aware, timestep-weighted objective. By propagating a reconstruction loss through the diffusion process, the \textit{DDPM-S} formulation explicitly prioritizes the physical integrity of the underlying pressure signal over a simple noise-matching schedule. This advancement allowed the model to outperform established deterministic baselines, achieving a 48\% reduction in \gls{MAE} compared to a standard \gls{MLP} and a 60\% reduction compared to an \gls{AE}+\gls{GPR} architecture. Crucially, this accuracy gain is concentrated in the physically most relevant regions: suction peaks, shock fronts, and control surface hinge lines, precisely the features whose degradation in deterministic surrogates directly corrupts integrated aerodynamic quantities such as wave drag and pitching moment.

The intrinsic stochasticity of the \gls{DDPM} was also exploited to provide a built-in assessment of prediction reliability. Beyond the statistical convergence of the ensemble quantities ($N \geq 50$), the sensitivity analysis showed that the sampling-induced spread remains spatially organized and concentrated in the most demanding aerodynamic regions, such as suction peaks, hinge-line discontinuities, and shock fronts. At the global level, the strong correlation observed between the \gls{GRI} and the reconstruction error (\gls{MAE}), with $\rho = 0.889$, indicates that operating conditions with larger spread also tend to be associated with larger surrogate errors in the present dataset. At the local level, the proposed \gls{LRI} curves show that regions with lower spread generally correspond to lower residuals, while high-spread regions are progressively associated with larger errors. Overall, this sensitivity analysis supports the interpretation of diffusion-sampling spread as a useful qualitative reliability diagnostic rather than as calibrated uncertainty quantification.

Taken together, these contributions build upon and extend diffusion-based surrogate studies in aerodynamics. While previous work demonstrated that \gls{DDPM} sampling variance provides qualitative variability maps 
for two-dimensional airfoil flows on structured grids \cite{liu2024uncertainty, ogbuagu2026foildiff}, the present framework 
advances this paradigm in three directions: (i)~the surrogate operates on a three-dimensional, unstructured surface mesh of $h = 139{,}374$ points, conditioned on a six-dimensional design space that includes control surface deflections; (ii)~it introduces a signal-aware training objective that improves reconstruction fidelity in sharp-gradient regions and yields a bounded, low-variance generative process (mean $\overline{\sigma}^{(N)} \approx 0.005$) that permits meaningful quantification of the residual stochastic spread; and (iii)~it formalizes the spread--error relationship through the \gls{LRI} and \gls{GRI}, which provide spatially resolved and global diagnostic measures of model sensitivity. These elements indicate that conditional diffusion surrogates are scalable to industrially relevant three-dimensional configurations and that their intrinsic stochasticity can serve as a practical reliability assessment tool.

This study is intentionally scoped to assess the value of a diffusion-based surrogate formulation for deterministic aerodynamic field prediction under limited-data conditions. The analysis is restricted to a single transonic CRM wing geometry, surface pressure as the target quantity, and a compact dataset of high-fidelity \gls{RANS} simulations. In addition, the benchmark is not intended to exhaust the full landscape of modern surrogate architectures for unstructured aerodynamic data, such as graph-based, operator-learning, or neural-field approaches. Within these boundaries, the present results indicate that conditional diffusion surrogates are a competitive and self-aware alternative for three-dimensional aerodynamic pressure prediction. Future work will explore the extension of this framework to full aircraft configurations, alternative target quantities such as skin friction, and the integration of the reliability diagnostics introduced here into active learning loops for adaptive \gls{CFD} database enrichment.

\section*{Acknowledgments}
This work has been supported by the TIFON project, ref. PLEC2023-010251/MCIN/AEI/ 10.13039/501100011033, funded by the Spanish State Research Agency. 

\bibliographystyle{elsarticle-num-names} 
\bibliography{sample.bib}

@article{catalani2024neuralfields,
  title     = {Neural fields for rapid aircraft aerodynamics simulations},
  author    = {Catalani, Giovanni and Agarwal, Siddhant and Bertrand, Xavier and Tost, Fr{\'e}d{\'e}ric and Bauerheim, Michael and Morlier, Joseph},
  journal   = {Scientific Reports},
  volume    = {14},
  pages     = {25496},
  year      = {2024},
  doi       = {10.1038/s41598-024-76983-w}
}

@article{shukla2024deepoperators,
  title     = {Deep neural operators as accurate surrogates for shape optimization},
  author    = {Shukla, Khemraj and Oommen, Vivek and Peyvan, Ahmad and Penwarden, Michael and Plewacki, Nicholas and Bravo, Luis and Ghoshal, Anindya and Kirby, Robert M. and Karniadakis, George Em},
  journal   = {Engineering Applications of Artificial Intelligence},
  volume    = {129},
  pages     = {107615},
  year      = {2024},
  doi       = {10.1016/j.engappai.2023.107615}
}

@article{hu2025pano,
  title     = {Physics-aware neural operator for high-fidelity fluid dynamics modeling with geometric and spectral priors},
  author    = {Hu, Xuebin and Ma, Qinglong and Zhao, Peizhi and Wang, Xun},
  journal   = {Physics of Fluids},
  volume    = {37},
  number    = {11},
  pages     = {115111},
  year      = {2025},
  doi       = {10.1063/5.0299765}
}

@article{xiang2025aerodit,
  author  = {Wang, Chunyang and Xiang, Hui and Jiang, Haoyu and Li, Bohan and Peng, Yutao and Fan, Zhenhua and Zhang, Miao and Li, Jianqiang},
  title   = {{AeroDiT}: Diffusion Transformers for {Reynolds-Averaged Navier--Stokes} Simulations of Airfoil Flows},
  journal = {Physics of Fluids},
  volume  = {37},
  number  = {12},
  pages   = {124120},
  year    = {2025},
  doi     = {10.1063/5.0256147},
}

@article{liu2025flowvitdiff,
  author  = {Liu, Yiming and others},
  title   = {A general framework for airfoil flow field reconstruction based on transformer-guided diffusion models},
  journal = {Chinese Journal of Aeronautics},
  year    = {2025},
  doi     = {10.1016/j.cja.2025.06.003},
  note    = {In press},
}

@inproceedings{he2017aerodatafusion,
  doi = {10.13009/EUCASS2017-108},
  author = {He, L. and Zhou, Y. and Qian, W. and Wang, Q.},
  title = {Aerodynamic Data Fusion with a Multi-fidelity Surrogate Modeling Method},
  booktitle = {Proceedings of the 7th European Conference for Aeronautics and Space Sciences},
  month ={3-6 july},
  address = {Milano,  Italy},
  year = {2017}
}

@inproceedings{anhichem2022multi,
  title = {Multifidelity Data Fusion Applied to Aircraft Wing Pressure Distribution},
  DOI = {10.2514/6.2022-3526},
  booktitle = {AIAA AVIATION 2022 Forum},
  publisher = {American Institute of Aeronautics and Astronautics},
  author = {Anhichem, M. and Timme, S. and Castagna, J. and Peace, A. and Maina, M.},
  year = {2022},
  pages={},
  month = {jun} 
}

@techreport{malik2012roleof,
  author      = {Malik, M. R. and Bushnell, D. M.},
  title       = {Role of Computational Fluid Dynamics and Wind Tunnels in Aeronautics R and D},
  institution = {National Aeronautics and Space Administration},
  number      = {NASA/TP--2012--217602},
  year        = {2012},
}

@article{yondo2019review,
  title={A review of surrogate modeling techniques for aerodynamic analysis and optimization: current limitations and future challenges in industry},
  author={Yondo, R. and Bobrowski, K. and Andr{\'e}s, E. and Valero, E.},
  journal={Advances in evolutionary and deterministic methods for design, optimization and control in engineering and sciences},
  doi = {10.1016/j.paerosci.2017.11.003},
  pages={19--33},
  year={2019},
  publisher={Springer}
}

@article{du2021rapid,
  title={Rapid airfoil design optimization via neural networks-based parameterization and surrogate modeling},
  author={Du, X. and He, P. and Martins, J. R.R.A.},
  journal={Aerospace Science and Technology},
  volume={113},
  pages={106701},
  year={2021},
  publisher={Elsevier},
  doi = {10.1016/j.ast.2021.106701},
}

@article{du2022airfoil,
  title={Airfoil design and surrogate modeling for performance prediction based on deep learning method},
  author={Du, Q. and Liu, T. and Yang, L. and Li, L. and Zhang, D. and Xie, Y.},
  journal={Physics of Fluids},
  volume={34},
  number={1},
  year={2022},
  publisher={AIP Publishing},
  DOI = {10.1063/5.0075784},
}

@article{solera2024betavae,
  title={$\beta$-Variational autoencoders and transformers for reduced-order modelling of fluid flows},
  author={Solera-Rico, A. and {Sanmiguel Vila}, C. and G{\'o}mez-L{\'o}pez, M. and Wang, Y. and Almashjary, A. and Dawson, S.T.M. and Vinuesa, R.},
  journal={Nature Communications},
  volume={15},
  number={1},
  pages={1361},
  year={2024},
  publisher={Nature Publishing Group UK London}
}

@article{frances2024betavae,
  title = {Toward aerodynamic surrogate modeling based on $\beta$-variational autoencoders},
  volume = {36},
  DOI = {10.1063/5.0232644},
  number = {11},
  journal = {Physics of Fluids},
  publisher = {AIP Publishing},
  author = {Francés-Belda, V. and Solera-Rico, A. and Nieto-Centenero, J. and Andrés, E. and Sanmiguel Vila, C. and Castellanos, R.},
  year = {2024},
}

@article{nandal2023synergistic,
  title={A Synergistic Framework Leveraging Autoencoders and Generative Adversarial Networks for the Synthesis of Computational Fluid Dynamics Results in Aerofoil Aerodynamics},
  author={Nandal, T. and Fulara, V. and Singh, R. K.},
  journal={arXiv preprint arXiv:2305.18386},
  year={2023},
  doi = {10.48550/arXiv:2305.18386}
}

@article{wang2023cgan,
  author={Wang, Y. and Deng, L. and Wan, Y. and Yang, Z. and Yang, W. and Chen, C. and Zhao, D. and Wang, F. and Guo, Y.},
  journal={IEEE Transactions on Neural Networks and Learning Systems}, 
  title={An Intelligent Method for Predicting the Pressure Coefficient Curve of Airfoil-Based Conditional Generative Adversarial Networks}, 
  year={2023},
  volume={34},
  number={7},
  doi={10.1109/TNNLS.2021.3111911}
  }

@article{dickstein2015noneq,
  title={Deep Unsupervised Learning using Nonequilibrium Thermodynamics},
  author={Sohl-Dickstein, J. and Weiss, E. A. and Maheswaranathan, N. and Ganguli, S.},
  journal={arXiv preprint arXiv:1503.03585},
  year={2015},
  doi = {10.48550/arXiv.1503.03585}
}

@article{ho2020ddpm,
  title={Denoising Diffusion Probabilistic Models},
  author={Ho, J. and Jain, A. and Abbeel, P.},
  journal={arXiv preprint arXiv:2006.11239},
  year={2020},
  doi = {10.48550/arXiv.2006.11239}
}

@article{nichol2021improved,
  title={Improved Denoising Diffusion Probabilistic Models},
  author={Nichol, A. and Dhariwal, P.},
  journal={arXiv preprint arXiv:2102.09672},
  year={2021},
  doi = {10.48550/arXiv.2102.09672}
}

@article{nichol2021beatgan,
  title={Diffusion Models Beat GANs on Image Synthesis},
  author={Nichol, A. and Dhariwal, P.},
  journal={arXiv preprint arXiv.2105.05233},
  year={2021},
  doi = {10.48550/arXiv.2105.05233}
}

@article{song2020score,
  title={Score-based generative modeling through stochastic differential equations},
  author={Song, Y. and Sohl-Dickstein, J. and Kingma, D. P. and Kumar, A. and Ermon, S. and Poole, B.},
  journal={arXiv preprint arXiv:2011.13456},
  year={2020}
}

@article{li2025research,
author = {Li, X. and Peng, Y. and Zheng, H.},
year = {2025},
title = {Research and Analysis of VAE, GAN, and Diffusion Generation Models},
volume = {1},
journal = {Science and Technology of Engineering, Chemistry and Environmental Protection},
doi = {10.61173/mr4pwd16}
}

@article{saharia2021iterative,
  title={Image Super-Resolution via Iterative Refinement},
  author={Saharia, C. and Ho, J. and Chan, W. and Salimans, T. and Fleet, D. J. and Norouzi, M.},
  journal={arXiv preprint arXiv:2104.07636},
  year={2021},
  doi = {10.48550/arXiv.2104.07636}
}

@article{lino2025diffusiongraph,
  title={Learning Distributions of Complex Fluid Simulations with Diffusion Graph Networks},
  author={Lino, M. and Pfaff, T. and Thuerey, N.},
  journal={arXiv preprint arXiv.2504.02843},
  year={2025},
  doi = {10.48550/arXiv.2504.02843}
}

@article{song2022ddim,
  doi = {10.48550/arXiv:2010.02502},
  author = {Song, J. and Meng, C. and Ermon, S.},
  title = {Denoising Diffusion Implicit Models},
  publisher = {arXiv},
  year = {2020},
  journal={arXiv preprint arXiv:2010.02502},
}

@article{kim2025pointwise,
  title={Point-wise Diffusion Models for Physical Systems with Shape Variations: Application to Spatio-temporal and Large-scale system},
  author={Kim, J. and Yang, S. and Kang, N.},
  journal={arXiv preprint arXiv:2508.01230},
  year={2025},
  doi = {10.48550/arXiv.2508.01230}
}

@article{ogbuagu2026foildiff,
title = {FoilDiff: A hybrid diffusion transformer model for airfoil flow field prediction},
journal = {Aerospace Science and Technology},
volume = {172},
pages = {111677},
year = {2026},
doi = {https://doi.org/10.1016/j.ast.2026.111677},
author = {Ogbuagu, K. and Maleki, S. and Bruni, G. and Krishnababu, S},
}

@article{liu2024uncertainty,
author = {Liu, Q. and Thuerey, N.},
title = {Uncertainty-Aware Surrogate Models for Airfoil Flow Simulations with Denoising Diffusion Probabilistic Models},
journal = {AIAA Journal},
volume = {62},
number = {8},
pages = {2912-2933},
year = {2024},
doi = {10.2514/1.J063440},
}

@article{xu2025diffpcno,
  title={Physically consistent and uncertainty-aware learning of spatiotemporal dynamics},
  author = {Xu, Q. and Bamber, J. L. and Thuerey, N. and Boers, N. and Bates, P. and Camps-Valls, G. and Shi, Y. and Zhu, X. X.},
  journal={arXiv preprint arXiv:2508.01230},
  year={2025},
  doi = {10.48550/arXiv:2508.01230}
}

@article{du2024turbulence,
  title = {Conditional neural field latent diffusion model for generating spatiotemporal turbulence},
  volume = {15},
  doi = {10.1038/s41467-024-54712-1},
  number = {1},
  journal = {Nature Communications},
  publisher = {Springer Science and Business Media LLC},
  author = {Du, P. and Parikh, M. H. and Fan, X. and Liu, X.-Y. and Wang, J.-X.},
  year = {2024}
}

@inproceedings{wei2024diffairfoil,
  title = {DiffAirfoil: An Efficient Novel Airfoil Sampler Based on Latent Space Diffusion Model for Aerodynamic Shape Optimization},
  doi = {10.2514/6.2024-3755},
  booktitle = {AIAA AVIATION FORUM AND ASCEND 2024},
  publisher = {American Institute of Aeronautics and Astronautics},
  author = {Wei, Z. and Dufour, E. R. and Pelletier, C. and Fua, P. and Bauerheim, M.},
  year = {2024}, 
}

@misc{kuntz2020markov,
  doi = {10.48550/arxiv.2001.02183},
  author = {Kuntz, J.},
  title = {Markov chains revisited},
  journal={arXiv preprint arXiv:2001.02183},
  year = {2020}
}

@article{guo2025noise,
  doi = {10.48550/arXiv.2502.04669},
  author = {Guo, Z. and Lang, J. and Huang, S. and Gao, Y. and Ding, X.},
  title = {A Comprehensive Review on Noise Control of Diffusion Model},
  publisher = {arXiv},
  year = {2025},
  journal={arXiv preprint arXiv:2502.04669}
}

@article{ronneberger2015unet,
  title={U-Net: Convolutional Networks for Biomedical Image Segmentation},
  author={Ronneberger, O. and Fischer, P. and Brox, T.},
  journal={arXiv preprint arXiv:1505.04597},
  year={2021},
  doi = {10.48550/arXiv.1505.04597}
}

@misc{zhan2024survey,
doi = {10.48550/arXiv.2409.19365},
author = {Zhan, Z. and Chen, D. and Mei, J.-P. and Zhao, Z. and Chen, J. and Chen, C. and Lyu, S. and Wang, C.},
title = {Conditional Image Synthesis with Diffusion Models: A Survey},
year = {2024},
journal={arXiv preprint arXiv:2409.19365}
}

@article{vaswani2017attention,
  author       = {Vaswani, A. and Shazeer, N. and Parmar, N. and Uszkoreit, J. and Jones, L. and Gomez, A. N. and Kaiser, L. and Polosukhin, I.},
  title        = {Attention Is All You Need},
  year         = {2017},
  doi={10.48550/arXiv:1706.03762}, 
  journal={arXiv preprint arXiv:1706.03762},
}

@inproceedings{vassberg2008crm,
  title = {Development of a Common Research Model for Applied CFD Validation Studies},
  DOI = {10.2514/6.2008-6919},
  booktitle = {26th AIAA Applied Aerodynamics Conference},
  publisher = {American Institute of Aeronautics and Astronautics},
  author = {Vassberg, J. C. and Dehaan, M. A. and Rivers, S. M. and Wahls, R. A.},
  year = {2008}, 
}

@article{sabater2022ddbb,
author = {Sabater, C. and St\"{u}rmer, P. and Bekemeyer, P.},
title = {Fast Predictions of Aircraft Aerodynamics Using Deep-Learning Techniques},
journal = {AIAA Journal},
volume = {60},
number = {9},
pages = {5249-5261},
year = {2022},
doi = {10.2514/1.J061234},
}

@article{catalani2026towards,
title = {Towards scalable surrogate models based on neural fields for large scale aerodynamic simulations},
journal = {Computers \& Fluids},
volume = {306},
pages = {106929},
year = {2026},
doi = {https://doi.org/10.1016/j.compfluid.2025.106929},
author = {Catalani, G. and Fesquet, J. and Bertrand, X. and Tost, F. and Bauerheim, M. and Morlier, J.},
}

@article{hines2023graph,
title = {Graph neural networks for the prediction of aircraft surface pressure distributions},
journal = {Aerospace Science and Technology},
volume = {137},
pages = {108268},
year = {2023},
doi = {https://doi.org/10.1016/j.ast.2023.108268},
author = {Hines, D. and Bekemeyer, P.},
}

@article{allmaras2012sa,
author = {Allmaras, S. and Johnson, F. and Spalart, P.},
year = {2012},
pages = {1-11},
title = {Modifications and clarifications for the implementation of the Spalart-Allmaras turbulence model},
journal = {Seventh International Conference on Computational Fluid Dynamics (ICCFD7)}
}

@inbook{kroll2014tau,
    author = {N. Kroll and S. Langer and A. Schw\"oppe},
    title = {The DLR Flow Solver TAU - Status and Recent Algorithmic Developments},
    booktitle = {52nd Aerospace Sciences Meeting},
    publisher = {American Institute of Aeronautics and Astronautics},
    year={2014},
    doi = {10.2514/6.2014-0080}
}

@article{tran2024electric,
  title = {Aerodynamics-guided machine learning for design optimization of electric vehicles},
  volume = {3},
  doi = {10.1038/s44172-024-00322-0},
  number = {1},
  journal = {Communications Engineering},
  publisher = {Springer Science and Business Media LLC},
  author = {Tran, J. and Fukami, K. and Inada, K. and Umehara, D. and Ono, Y. and Ogawa, K. and Taira, K.},
  year = {2024},
}

@article{ba2016layernorm,
  doi = {10.48550/arXiv.1607.06450},
  author = {Ba, J. L. and Kiros, J. R. and Hinton, G. E.},
  title = {Layer Normalization},
  year = {2016},
  journal={arXiv preprint arXiv:1607.06450},
  
}

@article{kingma2014adam,
author = {Kingma, D. and Ba, J.},
year = {2014},
month = {12},
pages = {},
title = {Adam: A Method for Stochastic Optimization},
journal = {International Conference on Learning Representations},
doi = {10.48550/ARXIV.1412.6980},
}

@article{smith2017super,
author = {Smith, L. N. and Topin, N.},
year = {2017},
month = {},
pages = {},
title = {Super-Convergence: Very Fast Training of Neural Networks Using Large Learning Rates},
journal = {arXiv preprint arXiv:1708.07120},
doi = {10.48550/arXiv.1708.07120},
}

@article{hawkins2003overfitting,
  title = {The Problem of Overfitting},
  volume = {44},
  DOI = {10.1021/ci0342472},
  number = {1},
  journal = {Journal of Chemical Information and Computer Sciences},
  publisher = {American Chemical Society (ACS)},
  author = {Hawkins, D. M.},
  year = {2003},
  month = dec,
  pages = {1–12}
}

@article{ying2019overfitting,
  title = {An Overview of Overfitting and its Solutions},
  volume = {1168},
  DOI = {10.1088/1742-6596/1168/2/022022},
  journal = {Journal of Physics: Conference Series},
  publisher = {IOP Publishing},
  author = {Ying, X.},
  year = {2019},
  month = feb,
  pages = {022022}
}

\end{document}